\documentclass{ar-1col-S2O}

\usepackage{hyperref}

\usepackage[comma]{natbib}
\usepackage{url}
\usepackage{arydshln}
\usepackage{amssymb, mhchem, wasysym}
\usepackage{gensymb}
\usepackage[T1]{fontenc} 

\usepackage[toc,page,titletoc]{appendix}
\usepackage[capitalise,nameinlink]{cleveref}

\jname{Xxxx. Xxx. Xxx. Xxx.}
\jvol{AA}
\jyear{YYYY}
\doi{10.1146/((please add article doi))}

\begin{document}

\markboth{Bergin et al.}{Carbon}

\title{Carbon from Interstellar Clouds to Habitable Worlds}

\author{Edwin A. Bergin,$^1$ Marc M. Hirschmann,$^2$ and Andr\'e Izidoro$^3$
\affil{$^1$Department of Astronomy, University of Michigan,  Ann Arbor, MI, USA, 48104; email: ebergin@umich.edu}
\affil{$^2$Department of Earth and Environmental Sciences, University of Minnesota, Minneapolis, MN, USA, 55414}
\affil{$^3$Department of Earth, Environmental and Planetary Sciences, Rice University, Houston, TX, USA, 77005}}

\begin{abstract}
Carbon is an essential element for a habitable world.   Inner (r$<$3~au) disk planetary carbon compositions are strongly influenced by supply and survival of carbonaceous solids.   Here we trace the journey of carbon from the interstellar medium to the processes leading to planet formation. The review highlights the following central aspects:

\hangindent=.3cm$\bullet$ Organics forming in evolved star envelopes are supplemented by aromatic molecules forming in the dense ISM to represent the seeds of (hydro)carbon supply through pervasive pebble drift to rocky planets and sub-Neptune cores. 

\hangindent=.3cm$\bullet$ Within the protoplanetary disk the sharp gradient in the C/Si content of Solar System bodies and mineral geochemistry outlines a tale of carbon loss from pebbles to within planetesimals and planets, and from planetary atmospheres.  

\hangindent=.3cm$\bullet$ Within two planet formation paradigms (pebble and planetesimal accretion) a range of planetary carbon content is possible that is strongly influenced by early ($<$ 0.5~Myr) formation of a pressure bump that titrates drift. 

Overall, it is unlikely that the carbon architecture of our Solar System applies to all systems.  In the absence of giant planets, carbon-rich rocky worlds and sub-Neptunes may be common.  We outline observations that support their presence and discuss  habitability of terrestrial worlds.
\end{abstract}

\begin{keywords}
carbon, exoplanets, solar system, planet formation, astrochemistry, geochemistry 
\end{keywords}
\maketitle

\tableofcontents

\newcommand\nodata{ ~$\cdots$~ }
\newcommand{\cc}{\mbox{cm$^{-3}$}}
\newcommand{\tauv}{\mbox{$\tau_V$}}
\newcommand{\av}{\mbox{A$_{\rm V}$}}
\newcommand{\ra}{\mbox{$\rightarrow$}}
\newcommand{\nhtwo}{\mbox   {n$_{\rm H_{2}}$}}
\newcommand{\nh}{\mbox {n$_{\rm H}$}}
\newcommand\ion[2]{#1$\;${\scshape{#2}}}%
\newcommand{\nat}{Nature}
\newcommand{\apj}{ApJ}
\newcommand{\aj}{AJ}
\newcommand{\araa}{ARA\&A}
\newcommand{\planss}{Plan. \& Space Sci.}
\newcommand{\apjl}{ApJ}
\newcommand{\apjs}{ApJS}
\newcommand{\aap}{A\&A}
\newcommand{\mnras}{MNRAS}
\newcommand{\icarus}{Icarus}
\newcommand{\ssr}{Space Sci. Rev.}
\newcommand{\prl}{Phys. Rev. Lett.}
\newcommand{\aapr}{A\&A Rev}
\newcommand{\pasj}{PASJ}
\newcommand{\maps}{Meteoritics \& Planetary Sci.}
\newcommand{\physrep}{Physics Reports}
\newcommand{\jcp}{J. Chem. Phys}
\newcommand{\apss}{{\it Astrophys.~Space~Sci.}}
\newcommand{\gca}{{\it Goechim.~Cosmochim.~Acta}}
\newcommand{\grl}{{\it Geophys.~Res.~Letters}}
\newcommand{\jgrp}{{\it J.~Geophys.~Res.~(Planets)}}

\newcommand{\pasp}{{\it PASP}}
\newcommand{\pnas}{{\it PNAS}}
\newcommand{\psj}{{\it Plan.~Sci.~J.}}
\section{INTRODUCTION} 

Carbon, the fourth most abundant element, is essential for our {\em living} planet.  With four free bonding sites, and a strong double bond, the chemistry of this volatile\footnote{In planetary science, volatility indicates how easily an element vaporizes from condensed phases. From the perspective of planet formation, highly volatile elements (C, H, O, and N) occur in molecules that readily sublimate from condensed form to gas (e.g., water, methane, CO, CO$_2$, N$_2$). } element is central to all life on Earth.  On Earth, the planetary carbon cycle regulates long-term global climate \citep{2010E&PSL.298....1D}.  Beyond the Solar System, the chemistry of interstellar space is dominated by the organic chemistry of carbon owing to the same chemical properties that make carbon essential for life \citep{2000ARA&A..38..427E}.   
The role of carbon in planet formation is distinct from water, and the distribution of carbon among planets and exoplanets is likely highly variable. Variable carbon content within the interiors of rocky planets influences the physics and chemistry of their surfaces and their enshrouding atmospheres.

\begin{figure}
\centering
\includegraphics[width=1.0\linewidth]{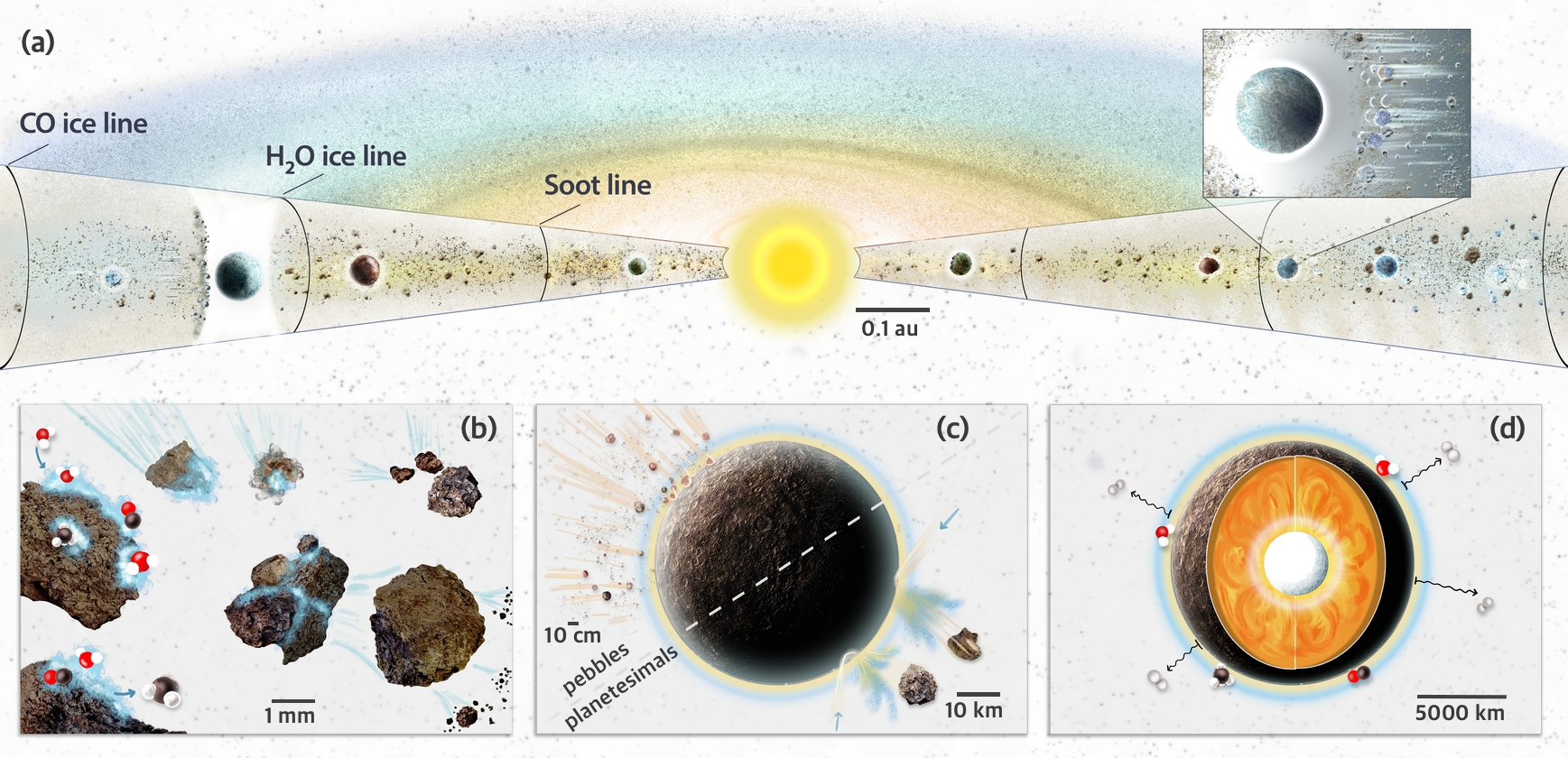} \caption{
Schematic of some of the key processes throughout planet formation that influence carbon supply and retention. (a) Key ice/grain sublimation fronts in the disk noted (CO$_2$ between water--CO and that of Silicates at inner disk edge not shown).  Planets forming in these locations will receive different compositions at birth \citep{Oberg11_C_O}. Dust that grows beyond the ``soot'' line \citep{Lodders2004, Kress2010, Li21} will be rich in refractory carbon content and at greater distances molecular ices. 
(blow-out:) The drift of carbon-rich pebbles from the outer disk is a key process in carbon supply.  Drift itself is influenced by pressure bumps that can lead to a pile-up of pebbles that can be potential sites for planetesimal and planet formation (shown on left side of disk).   (b) Solids accrete ice mantles, sublimate key components at sublimation fronts and can collide leading to both growth and fragmentation.  (c) Rocky planets and cores of gas-rich planets grow and gain carbon through accretion of pebbles (top) and/or planetesimals (bottom).  Impacts of larger planetesimals can also strip atmospheric gas and also supply material.  (d) heavy element differentiation can take carbon to the core, and the resulting magma ocean/atmosphere equilibrium can influence atmospheric composition with lighter gases escaping depending on the planet mass.
Figure credit to N. Fuller and Sayo Studio.
}
    \label{fig:megaschematic}
\end{figure}

In this review, we explore the present understanding of the distribution and exchange of carbon during stellar and planetary birth, drawing from methods and results in both the Earth and Astronomical sciences. Our primary focus will be planet formation in the inner disk where rocky planets and sub-Neptunes form. Fig.~\ref{fig:megaschematic} illustrates some of the key processes active in protoplanetary disks and in growing planets. 
  In \S 2 we discuss the evolutionary astrophysical constraints on carbon chemistry from diffuse clouds to forming planetary systems. In \S 3 we summarize carbon distribution in stars, Solar System bodies, polluted white dwarfs, and exoplanets.  We explore carbon supply and loss associated with planet formation for different accretion paradigms in \S~4; while \S 5 summarizes expectations for carbon in exoplanetary bodies. Finally, in \S 6 we discuss the impact of the planet carbon content on planetary habitability.
Several recent reviews in this area are also excellent sources of information \citep[e.g.,][]{2023ASPC..534..907L, 2023ASPC..534.1031K}.   

\subsection{Carbon and Water In Planet Formation} 

The fates and chemical roles of carbon and water during planetary formation are distinct.  In the interstellar medium (ISM) and disks, carbon chemistry is driven largely by nonequilibrium (kinetic) processes. In contrast, chemical reactions between water vapor and ice, the main reservoirs of H$_2$O, approach equilibrium more easily under a wide range of conditions \citep{vanDishoeck13}.

Within a planet-forming disk, the isotherm that governs ice/vapor equilibrium (the ``snow'' or ``ice'' line) is well-established as a central facet of planet formation \citep{Stevenson88, Ros13}. The backward diffusion of sublimated water across the ice line causes recondensation, promoting solid growth and water retention.  In contrast, carbon is supplied to (inner few au) planet-forming regions mostly in the form of refractory organics \citep{Bergin15, Alexander17} which undergo irreversible sublimation/pyrolysis at the ``soot'' line \citep{Li21}. Furthermore, as rocky planets are constructed, thermal processing causes differentiation between silicate melts, atmospheric vapor, and an Fe-rich alloy destined for the metallic core \citep{wood2013carbon}.  This leads to a strong redox dependence to the carbon distribution among the principal planetary reservoirs \citep{hirschmann2012magma,Gaillard22}.

\subsection{The Interstellar Conundrum} 
\label{sec:conundrum}

A fundamental feature of carbon mass transfer during planet formation is seen in the relative carbon content of the Earth compared to the materials provided by collapse which are thought to reflect the pre-stellar stage.  The paucity of water on Earth suggests the formation inside the water ice line near $\sim125$~K, so the planet was constructed from refractory materials initially found in dust grains.   
In ISM, almost all silicon is found in the solid state, and $\sim$ 50\% of carbon is found in the gas \citep{2021ApJS..257...63Z}. The remaining C is present as a population of refractory carbonaceous grains that are provided to the solar nebular disk through collapse.  Dividing the cosmic elemental abundance of carbon relative to silicon by two yields C/Si $\sim$5 for interstellar grains \citep{2021A&A...653A.141A, 2021A&A...656A.113A}.  Rocky planets constructed from such material should be carbon-dominated rock, but Earth and the other terrestrial planets in our Solar System are not.\footnote{This was first called out to EAB by the late Dr. Michael Jura, who remains an inspiration.}

\begin{figure}
    \centering
    \includegraphics[width=1.0\linewidth]{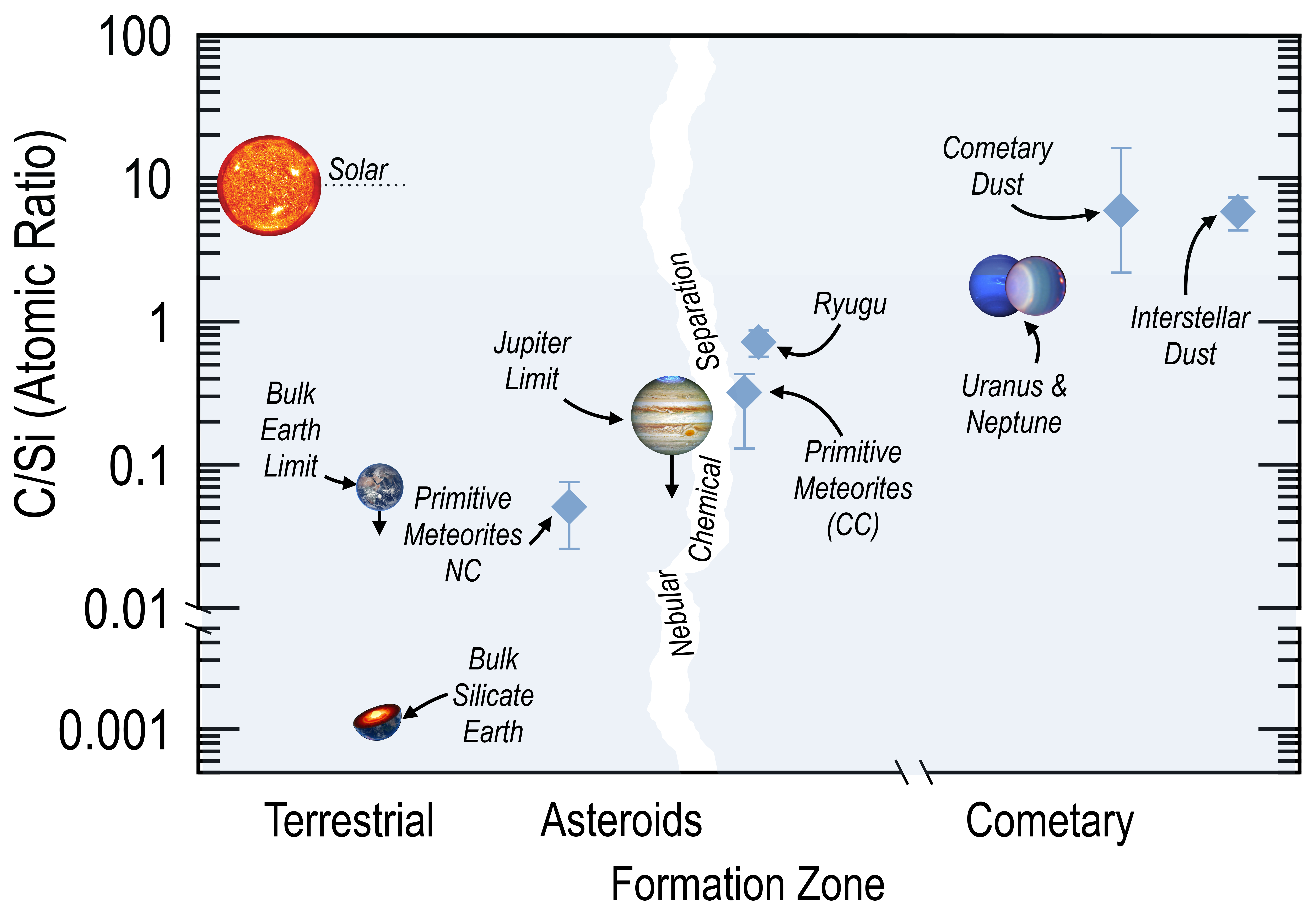}
    \caption{Bulk C/Si (atomic ratio) of bulk in Solar System bodies in the relative context of their potential formation location.  There is a noted chemical separation of the early ($<$1~Myr) Solar System, that is associated with the formation of Jupiter \citep{Kruijer20}.  In this figure we denote this as ``nebular chemical separation'', but this has some uncertainty in location.  
    The solar value uses the estimates obtained by \citet[][Si]{2021A&A...653A.141A} and \citet[][C]{2021A&A...656A.113A}.
    The bulk Earth upper limit is taken from \citet{Li21}.  Bulk Silicate Earth and the C/Si ratios for NC and CC chondrites are from \citet{Bergin15}.  For Jupiter see \S~\ref{sec:Jovian}.
    The value from Hyabusa/Ryugu is given by \citet{2022PJAB...98..227N} and the value for cometary dust is an average of comets Halley and 67P \citep{Bergin15, Rubin19}.  The C/Si ratio for Uranus and Neptune is based on the work of \citet{2024Icar..42116217M}.
    Error estimates for the Sun, BSE, Uranus and Neptune are smaller than the symbols. Error bars for NC and CC represent the range of values within existing samples while all other values are estimated measurement errors.
    }
    \label{fig:C_Si_solarsys}
\end{figure}

To illustrate this conundrum graphically, in Fig.~\ref{fig:C_Si_solarsys} we plot atomic C/Si ratios for Solar System bodies along with the carbon content of interstellar dust \citep[see also][]{1987A&A...187..859G, Lee10, 2014prpl.conf..363P}.  The Bulk Silicate Earth (BSE) is $\sim$4 orders of magnitude depleted compared to the Sun and interstellar dust.  A portion of this carbon may be sequestered in the Earth's core, but a generous upper limit \citep{Li21} still renders the Bulk Earth to be highly carbon-depleted compared to primitive materials (Fig.~\ref{fig:C_Si_solarsys}). Exploring other Solar System objects, chondritic meteorites, both noncarbonaceous (NC) and carbonaceous (CC), and asteroids Ryugu (Hyabusa) and Bennu (Osiris-Rex) are carbon-poor while comets Halley and 67P are carbon-rich, with carbon concentrations comparable to interstellar dust.  

\begin{marginnote}[]
\entry{Bulk Silicate Earth}{(BSE) is the composition of the Earth, excepting that fraction sequestered in the metallic core.}
\entry{Bulk Earth}{The Bulk Earth is the composition of the entire Earth, including the core.}
\entry{NC}{Non-Carbonaceous Chondrites are primitive meteorites that include ordinary and enstatite chondrites and are comparatively poor in volatiles.}
\entry{CC}{Carbonaceous Chondrites are primitive meteorites, including the rare CI chondrites, that are rich in carbon and water.}
\entry{Soot line}{Theorized location in a disk where carbonaceous solids undergo irreversible sublimation.}
\end{marginnote}

\subsection{The Forms of Carbon, Isotopes, and Nomenclature} 
\label{sec:carbonforms}

In our review we focus on the forms of carbon that are most likely to be supplied to  planet interiors (in exo-Earths, super-Earths, and sub-Neptunes) which form in the inner (r $<$ 3 au for a solar mass star) regions of the planet-forming disk.  This places the focus on solid state carbon carriers.  
Carbonaceous solids inferred to be present in the ISM and seen in primitive solar system materials (e.g. meteorites, asteroids, comets) often, but not always, involve bonds with hydrogen where there are two main classes, aliphatic and aromatic. These hydrocarbons (and other species with O and N) can be combined into a lattice structure (i.e. a grain).  For simplicity, these are defined below.

\vspace{-5mm}
\begin{extract}
{\bf aliphatic:} Carbon bonded in straight or branched chains (also cyclic).  Electrons are shared only between adjacent carbon atoms.  Methane (CH$_4$) and propane (C$_3$H$_8$) are key examples.

{\bf aromatic:} Cyclical carbonaceous compounds characterized with three $\sigma$ bonds where electrons are concentrated between the bonding atoms.  The fourth electron of the carbon is in a p-orbital that is shared with all other C atoms in the ring.  Benzene (C$_6$H$_6$) is a common example, as is naphthalene (C$_{10}$H$_{8}$) the simplest polycyclic aromatic hydrocarbon.
\end{extract}

Information on the form of solid state carbon comes from models of interstellar grains (\S~\ref{sec:diffuseism}), observations of interstellar ices (\S~\ref{sec:denseism}), and from detailed studies of primitive solar system solids (\S~\ref{sec:smallbodies}).  More detailed aspects of these various components are discussed in this review. Here, for clarity, we introduce key aspects with the various forms listed in terms of volatility; the carriers with the lowest volatility represent the highest likelihood of being incorporated by a forming planet. In the Supplementary Materials we also include information on isotopic ratios measured within each component.  These ratios provide information on the physical conditions at formation, in particular formation at T $<$ 70~K.

\begin{extract}
{\bf Carbon-Oxygen ices:} CO and CO$_2$ are abundant carbon carriers in the dense regions of the ISM, but in the inner disk are found in the gaseous state given their low sublimation temperatures \citep[CO:$\sim$25~K, CO$_2$: $\sim$45~K, for relevant pressures;][]{2022ESC.....6..597M}.

{\bf Soluble Organic Matter (SOM):} Soluble organics are defined by study of meteoritic (or asteroid sample return) material and represent the fraction of carbon solids that can be extracted using polar solvents. In carbonaceous chondrites (a class of primitive meteorites), carbon carriers in the SOM represent a wide diversity of compounds, including sugars, amino acids, and polycyclic aromatic hydrocarbons \citep[][see \S~\ref{sec:smallbodies}]{sephton2002organic}.  The diverse chemical complexity in SOM is generated by the energetic processing of  simpler ices and also by the action of liquid water  during the phase in which the $^{26}$Al decay was active \citep{Lee2025}.   Given the genetic link to simpler ices, we also classify (less complex) interstellar organic ices (e.g. CH$_3$OH) in this category. Many of the simple interstellar ices sublimate with water near $\sim$150~K and there is a close correspondence between the composition of interstellar and cometary ices \citep{2000A&A...353.1101B}.

 {\bf  Polycyclic Aromatic Hydrocarbons (PAHs):} Molecules consisting of two or more fused carbon rings.  Many PAHs have planar structures, but some more complex forms (e.g. Coronene) can deviate from this structure.
PAHs are readily observed in interstellar space and are key carbon carriers in solids from the solar system and in dense ISM \citep{Tielens08}.
Estimates suggest that most interstellar PAHs contain $\sim$40--55 carbon atoms \citep{2023FaDi..245..380L}.
PAHs are, in general, less volatile than water ice \citep{doi:10.1021/je7005133}.

{\bf Insoluble Organic Matter (IOM):} The majority of carbon carried  within primitive bodies, both asteroids and comets, appears to be found in a more refractory macromolecular form that is insoluble when exposed to typical solvents \citep{Alexander17, Fray2016, Rubin19}.  Detailed laboratory investigation of the IOM in chondritic meteorites finds that it is composed of small aromatics (1--4 rings) cross-linked by aliphatic branches creating a molecular structure, i.e. generally smaller than interstellar PAHs \citep{1977GeCoA..41.1325H, Derenne2010}.  This material is similar to terrestrial kerogen, which is insoluble in normal solvents \citep{1987GeCoA..51.2527K}; although there are chemical differences~\citep{Alexander17}.    The origin of the IOM has some uncertainty but could originate in the dense ISM and/or the outer disk through energetic (UV, X-ray, Cosmic-Ray, e$^-$) processing of interstellar ices (\S~\ref{sec:denseism}, \ref{sec:disk}, \& \ref{sec:smallbodies}).  
Similar to soluble organics, the IOM has been altered by parent body processing including aqueous alteration transforming a portion of the SOM into the IOM or perhaps even operating in the other direction \citep{LeGuillou2014}.
Cometary bodies also have a significant macromolecular organic content which is analogous to the IOM \citep[sometimes called CHON particles;][]{1999SSRv...90..109F,Fray2016, 2019A&A...630A..27I}. 
\end{extract}

\section{THE ARC OF CARBON FROM ATOMS TO MACROMOLECULES}
\label{sec:assay}

The story of carbon in planet formation begins with grains formed in the diffuse and dense ISM that are supplied to the disk and ends with planetary differentiation. Although the initial stages are dominated by gains, the early and late stages of planetary birth are dominated by condensed carbon loss. A list of key mechanisms active in the gain and loss of condensed carbon and the size scale of the solids on which they operate are given in  Table 1 of Supplementary Materials.

In Table~\ref{tab:c_assay} we provide an assay of the carbon content and major carriers through the various stages of star and planet formation, which is explored in this section in light of  key processes given in Supplementary Table~1.   Table~\ref{tab:c_assay}  illustrates the presence of two carbon pools: one more refractory and one more volatile.  During the various phases, the volatile component cycles between gas and ice depending on condensation and sublimation, but can be transformed into more or less volatile forms through gas phase or grain surface chemistry.  The refractory component appears to be (or is assumed to be) more constant across evolutionary stages, but this hides significant loss and inferred production.
This is discussed in greater detail below.

\subsection{Diffuse ISM}
\label{sec:diffuseism}

\begin{textbox}[h]\section{DIFFUSE ISM EVOLUTION}
The dynamic multiphase interstellar medium is fed by stellar mass loss and is shaped by supernova explosions.  The diffuse ISM  forms via the influence of large scale processes \citep[see][]{2020SSRv..216...76B} such as supernova explosions, stellar winds, and/or
colliding turbulent flows with sufficient ram pressure to trigger thermal instability within the warm (T$_{\rm gas} \sim$ 1000~K) neutral medium producing cooling and compression of the gas to higher densities (\nh\ $\sim$ 1 $\rightarrow$ $10^{2\text{--}3}$~\cc ) and lower temperatures (T$_{\rm gas} \sim$ 20-50~K).  
The grains are composed of silicate and carbonaceous forms characterized by a steep size distribution with an average size of $\sim$0.1~$\mu$m \citep{Mathis77}. 
Carbonaceous dust is known to form via combustion and condensation within the outflowing envelopes of Asymptotic Giant Branch (AGB) stars, but it can also be produced in supernovae \citep{1989ApJ...341..372F, 1992ApJ...401..269C,1988A&A...206..153G, Jager09,2024A&ARv..32....2S}.  
Interstellar shocks are suggested to be a primary destruction mechanism for grains \citep{2014A&A...570A..32B}.  A mismatch between the formation rates and the destruction rates of carbonaceous grains suggests that dense ISM may also contribute to the formation of carbon grains.
\end{textbox}

Carbon forms deep in stellar interiors via He fusion \citep[i.e., 3$\alpha$ process;][]{1957RvMP...29..547B} and is seeded into space as refractory solids and in atomic form via mass loss during the later phases of stellar evolution.
In the neutral phases of diffuse galactic gas (n$_{\rm H} < 100$~\cc, \av $<$ 0.2$^m$), \ion{C}{ii} is the primary carrier (see the sidebar titled: ``Diffuse ISM Evolution'' for additional context).
  The remainder is found in dust grains, where its presence is inferred by a variety of means.
Observations of atomic lines (e.g. \ion{C}{ii} 2324.4 \AA ) in absorption towards diffuse clouds (\av\ $\sim 1-2^m$) find $\sim$50\% of the carbon missing from the gas phase \citep{1996ARA&A..34..279S}.  Correlatively, there are several features attributed to carbonaceous components in interstellar dust, suggesting that the remaining carbon is in the solid state.  These include a prominent feature in the extinction curve, i.e., the ``bump'' seen at 2175~\AA\  and features associated with aliphatics (3.4 $\mu$m) and aromatics (3.3, 6.2, 7.7, 8.6, 11.2, and 12.7 $\mu$m; these are labeled Aromatic Infrared Bands or AIB) in the mid-infrared \citep{1998Sci...282.2204H}. There are also additional absorption bands associated with hydrocarbon dust \citep{2013ApJ...770...78C, 2025ApJ...992....8P}.
A more complete discussion of carbon carriers, including, e.g. diffuse interstellar bands, can be found in the review by \citet{1998Sci...282.2204H}.

 Observations of IR features associated with hydrocarbon dust show evidence of aliphatic and aromatic compounds with an evolution towards aromatic compounds with increased UV exposure \citep{2007ApJ...662..389G,2008A&A...490..665P}.
   Models of overall dust extinction and emission generally include silicates along with some mixture of organic material which is assumed to be separate or as a single composite \citep{2013ApJ...770...78C,2023ApJ...948...55H, 2024A&A...684A..34Y}.   Initial interstellar dust models focus on graphite as a key carbon carrier \citep[e.g.,][]{1984ApJ...285...89D}, but modern models (references above) typically assume that grains are composed of some form of aliphatic and aromatic hydrocarbons in an amorphous form with a range of hydrogen content (labeled as a:C-H).  Some models have amorphous hydrocarbons that coat the silicate grains in a manner that is reminiscent of some materials from the solar system \citep{flynn2013organic}.

\begin{figure}
    \centering
    \includegraphics[width=0.9\linewidth]{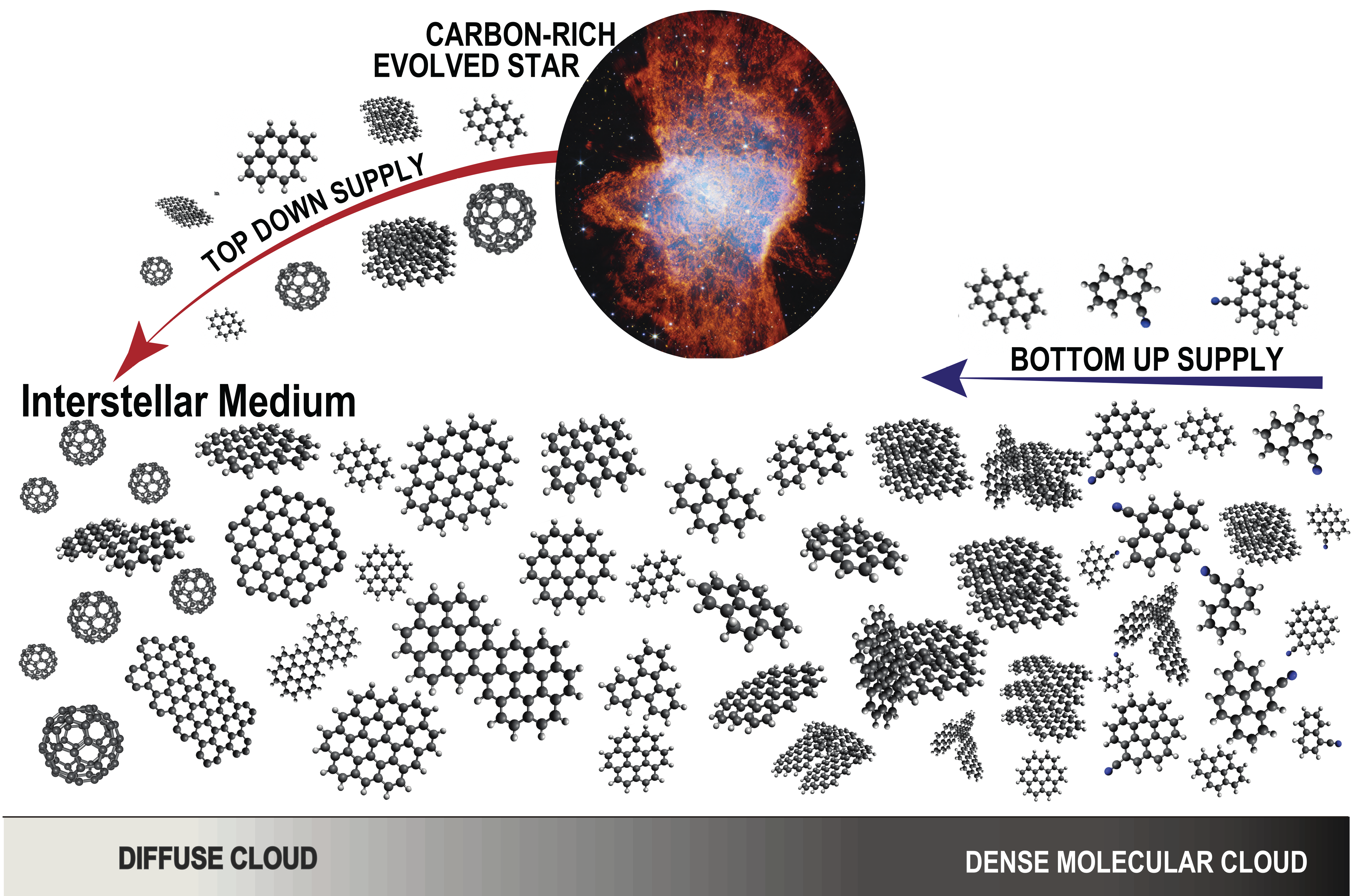}
    \caption{Schematic of the evolution of carbonaceous grains in the diffuse to dense ISM.  Carbon-rich grains form in evolved stars and are provided the diffuse interstellar medium (labeled as top down). These grains can be destroyed in shocks and processed by radiation such that only the stablest forms are provided to the dense cloud.  In dense molecular clouds there is the possibility of bottom up supply of small PAHs that could lead to grain evolution subsequent phases.  This figure predominantly illustrates the evolution of aromatic as opposed to aliphatic carbon carriers which is discussed in the text.
    Figure adapted from \citet{2015ApJ...807...99A,2021EPJD...75..152Z} and kindly produced by A. Candian.
    }
    \label{fig:cgrains}
\end{figure}

  Various aspects of carbon dust formation are 
 shown schematically in Fig.~\ref{fig:cgrains}. Carbon dust formation in AGB stars creates large hydrocarbon ring molecules that can lead to larger planar structures and also to the formation of fullerenes with cage-like structures, which are supplied to the ISM.
Given the high temperatures of formation ($\sim$1000~K)
these grains hold the isotopic signature of the local ISM at their time of birth.  
 The important physical and chemical aspects of this formation and grain nucleation are discussed by \citet{2011EAS....46..293J}.
 Depending on the gas density and magnitude of the UV field, smaller grains can be ionized, gain or lose hydrogen atoms, and in more exposed environments, leave  only the most stable species \citep[Fig.~\ref{fig:cgrains};][]{2016A&A...595A..23A}.  In general, interstellar dust is argued to be more aromatic in form as opposed to aliphatic \citep[e.g., 10$\times$ more aromatic material by mass;][]{2025ApJ...992....8P}; aliphatic material is more easily fragmented and subject to hydrogen loss \citep{2021A&A...652A..42M}.  
 The effective destruction of carbonaceous grains by shocks in the diffuse ISM suggests that there needs to be an additional bottom-up production mechanism, perhaps in the cold (T $\sim$10-30~K) and dense (n $\sim$10$^{4}$~cm$^{-3}$) ISM \citep{2014FaDi..168..313J, 2020ApJ...892...96R}. 
 The relation of ISM grains with the meteoritic IOM and the SOM remains unclear.   \citet{2017M&PS...52.1797A} explores this and finds that on the order of 10\% of primordial dust may have survived within primitive meteorites.  However, given the mismatch in destruction rates of carbon dust and production rates in AGB stars, a central question is whether any dust production in the dense ISM or protoplanetary disk is supplied back to the ISM - which would provide a different link between planetary system materials and interstellar dust.
 Overall, given the dynamic nature of the evolution of the ISM, it is likely that different molecular clouds, which themselves originate via large-scale shocks, are seeded with variable proportions of solid-state and gaseous carbon.

\begin{table}[h]
\tabcolsep7.5pt
\caption{Carbon Assay: Diffuse gas thorough Gas-Rich Protoplanetary Disks$^{\rm a, b}$}
\label{tab:c_assay}
\begin{center}
\begin{tabular}{|l|r|c|c|c|c|c|c|}
\hline
\multicolumn{1}{|c}{ISM Phase} & \multicolumn{2}{|c|}{Vapor} & \multicolumn{2}{c|}{Ice} & \multicolumn{2}{c|}{Refractory} & \multicolumn{1}{|c|}{Refs.$^{\rm c}$} \\ 
\multicolumn{1}{|c|}{} & \multicolumn{1}{c}{C/H} & \multicolumn{1}{c}{Carrier} & \multicolumn{1}{|c}{C/H} & \multicolumn{1}{c}{Carrier} & \multicolumn{1}{|c}{C/H} & \multicolumn{1}{c|}{Carrier} &\multicolumn{1}{|c|}{} \\ 
\hline
Diffuse ISM & & & &  & & & \\
 & \bf{107} & \bf{C~{\scriptsize II}} & \nodata & \nodata & \bf{107} & \emph{dust} & (1) \\\hline
Dense ISM &  & & & & &  &  \\
\hspace{0.15cm}(quiescent) & {\bf 49} & {\bf CO} & {\bf $\sim$50} & \bf{CO/CO$_2$}& \texttt{\textsl{$\le$111}} &  \textit{dust} & (2) \\
& \textbf{$\sim$1} & \textbf{H-rich org.} & \textit{$>$3} & \textit{PAHs}&   &   & \\\hdashline
\hspace{0.15cm}(star-form.) & \textsl{$\sim$97} & \textbf{CO} & \nodata & \nodata &  \texttt{\textsl{$\sim$106}} & \textit{dust} &    \\
 & \textbf{$\sim$3} & \textbf{O-rich org.} & \nodata & \nodata &  &  & (3) \\
 & \textbf{8} & \textbf{PAHs} & \nodata & \nodata &  &    &\\\hline
Disks &  & & & & & &  \\
\hspace{0.15cm}(r $>$10~au) & \textbf{10} & \textbf{CO} & \textbf{$>$13} & \textbf{CO/CO$_2$} &\textit{107} & \textit{dust} & (4)  \\
 & \textit{10} & \textit{H-rich org.} & & & \texttt{\textsl{$<$74}}& unknown  &  \\ \hdashline
 \hspace{0.15cm}(r $<$2~au) & \textit{10-102} & \textbf{CO} &\nodata&\nodata  &\textit{$\leq$107} & \textit{dust} &   \\
    & \textit{$\sim$10} & \textbf{CO$_2$} & \nodata & \nodata  & & & (5) \\
      & \textit{0 $-\sim$90} & \textbf{H-rich org.} & \nodata & \nodata & & &\\
\hline
\end{tabular}
\end{center}
\begin{tabnote}
$^{\rm a}$Normalized to a  total carbon content of 214ppm \citep{2012A&A...539A.143N} with an assumed water ice abundance of 100ppm for calculation of ice abundances given relative to water. $^{\rm b}$Quantities and/or carriers given in {\bf bold} font are known via direct observations, those given in \emph{italics} are inferred (i.e. indirectly known to be present via detailed models), while values given in \texttt{\textsl{slanted}} form are assumed. $^{\rm c}$References are: (1) \citet{2021ApJS..257...63Z}; 
(2) \citet{Xu_TMC_Census, 2023NatAs...7..431M, 2024Sci...386..810W, 2025ApJ...984L..36W}; (3) \citet{2014FaDi..168...61B, 2005pcim.book.....T}; (4) \citet{ 1994ApJ...421..615P, 2021ApJS..257....7B, 2024ApJ...975..166B}; (5) \citet{1994ApJ...421..615P,2024ApJ...977..173C, 2024A&A...689A.231K}.
\end{tabnote}
\end{table}

\begin{textbox}[h]\section{DENSE ISM EVOLUTION}
The formation of many pc-sized molecular clouds produces a highly structured filamentary medium with centrally concentrated cores (0.1~pc scale) condensing within the filaments.  Parsec-scale filamentary structures with typical central densities of $\sim 10^{4-5}$~cm$^{-3}$ and temperatures of $\sim$20~K \citep{2019A&A...621A..42A}.  Centrally concentrated ``pre-stellar cores'' have central densities that generally exceed 10$^5$ \cc with  T $\sim$ 10~K and are characterized by complex sub-sonic internal motions and rotation \citep{2007ARA&A..45..339B}.
The subsequent collapse of the rotating core leads to the disk formation  surrounded by a collapsing envelope that is ablated along the polar axes by energetic outflowing material entrained within winds and jets.  
Protostars with masses below $\sim$8~M$_\odot$ have their luminosity dominated by the release of gravitational potential energy via accretion \citep{2007ARA&A..45..565M}, which heats the forming disk and the surrounding infalling envelope.
\end{textbox}

\subsection{Dense ISM}
\label{sec:denseism}

We discuss the carbon inventory in the dense ISM through the lens of two phases (see the sidebar entitled: ``Dense ISM Evolution'' for additional context).   The first ``quiescent and cold molecular gas'' precedes the birth of the star with the formation of simple and complex\footnote{By interstellar standards the formation of molecules with $\ge$6 atoms is complex \citep{2009ARA&A..47..427H}.} molecules in the gas and as icy mantles coating grain surfaces.   The second is labeled ``star-forming and warm molecular gas''.  In this stage, the birth of the star heats the surrounding material sublimating both water and organic ice, revealing the organic richness of the interstellar inventory.
On the basis of decades of observations of molecular species, it is fair to state that carbon is the backbone of interstellar chemistry.  The story of star and planet birth is the formation of CO, molecular  hydrogen, organics, and water. This begins in the dense interstellar medium.

\begin{figure}
    \centering
    \includegraphics[width=0.9\linewidth]{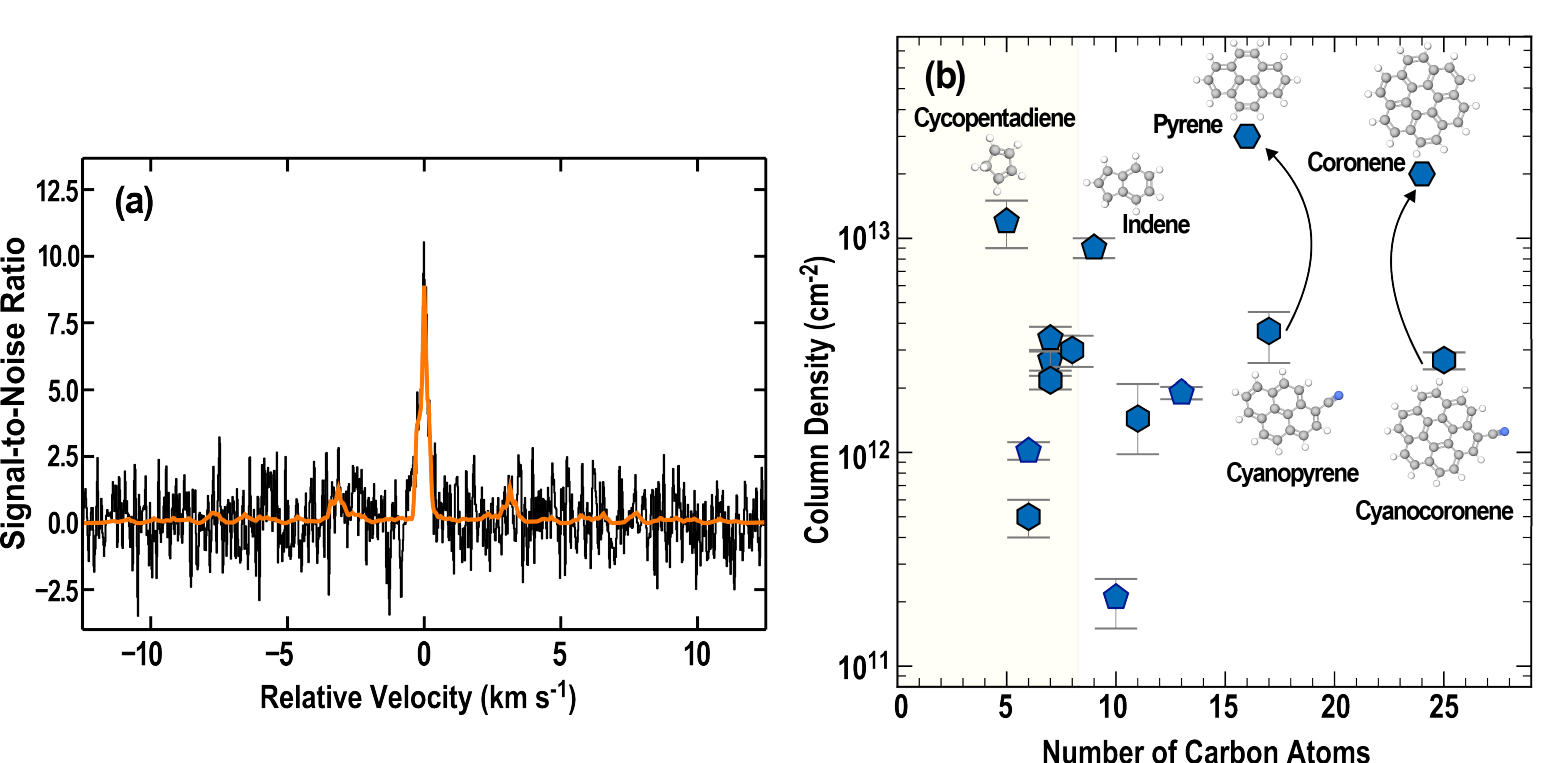}
    \caption{(a) Signal to noise ratio of the spectral line stacked detection of the 7-ring PAH cyanocoronene in a cold dark cloud.   (b) Detected molecular column density of complex carbon compounds in TMC-1.  Two symbols are shown: pentagons to reflect the presence of a 5-member carbon ring and hexagons to reflect the presence of more stable six-member carbon rings.  The yellow shaded area notes the regime of single ring cycles and beyond that point detected (or inferred) species contain multiple benzene and/or pentagon rings.  The arrows from cyanopyrene and cyanocoronene reflect the fact that the pyrene and coronene abundances are inferred from the detections of the cyano-substituted derivative \citep{2025NatAs...9..262W}.  In the molecular compounds the carbon atoms are gray, hydrogen as white, and nitrogen as blue.
    Figure adapted from \citet{2025ApJ...984L..36W}; the list of molecules shown in this plot is found in Table F1 of that paper. For conversion to molecular abundances the total column of molecular hydrogen along this line of sight is 1.8 $\times$10$^{22}$~cm$^{-2}$ \citep{Xu_TMC_Census}.   Creative Commons Attribution 4.0 License.
    }
    \label{fig:ismpah}
\end{figure}

\subsubsection{Quiescent and Cold Molecular Gas}
\label{sec:quiescientgas}

 A central example of the cold and quiescent dense medium is the chemically rich TMC-1 core \citep{2004PASJ...56...69K, Xu_TMC_Census} but extends towards numerous other objects \citep{2024MNRAS.533.4104S}.   In these clouds, it is a general assumption that $\sim$50\% of the carbon is locked in the refractory grain cores, but this number likely varies (\S~\ref{sec:diffuseism}).

  The most prominent gaseous carbon carrier  is CO with an abundance of $\sim$100 ppm relative to H.  However, the presence of increasing density and decreasing temperature gradients toward the center facilitates the deposition of gaseous molecules on grain surfaces \citep{2007ARA&A..45..339B}.  
  Catalytic reactions on grain surfaces, potentially aided by photoprocessing \citep[][]{2009ARA&A..47..427H}, transforms CO into less volatile forms principally CO$_2$ and oxygen-rich organics.   
  CO and CO$_2$ ice are observed through their vibrational movements and appear to comprise the most abundant forms of carbon ices \citep{2023NatAs...7..431M}.  Detections of more complex molecules embedded alongside water are difficult as the functional group stretch of, e.g. CH$_3$, could potentially be held by numerous organics \citep{Boogert15}.  However, CH$_3$OH ice is detected \citep{2023NatAs...7..431M}.  
  
  Although CO dominates the carbon inventory, another important facet of the quiescent gas composition is emerging. Dense molecular gas is associated with detection of highly unsaturated hydrocarbon chains (no rings) and  cyanopolynes  \citep[][and references therein]{2004PASJ...56...69K,2013ChRv..113.8981S,Xu_TMC_Census}. Two studies, \citet{2018Sci...359..202M} and \citet{2021A&A...649L..15C}, revolutionized our understanding of the chemistry of quiescent gas through the direct detection of large aromatic (PAH) molecules. 
  The AIB bands have long been associated with PAHs, but identification of individual carriers is challenging. This has been overcome in dense gas, possibly capturing the {\em in situ} transition from large molecules to small grains (Fig.~\ref{fig:cgrains}). Stable PAHs including benzene (1-ring), naphthalene (2-ring), pyrene (4-ring) and coronene (7-ring) are present in $\sim$10~K gas \citep{2018Sci...359..202M, 2021Sci...371.1265M, 2024Sci...386..810W, 2025ApJ...984L..36W}.

 The detection of 7-ring PAH cyanocoronene is shown in Fig.~\ref{fig:ismpah} together with the column densities of ringed molecules.    There is some uncertainty about the formation pathways and it could be from the ``bottom-up'' constructed through the predominant radical/ion-molecule chemistry that creates less complex species \citep[][]{2009ARA&A..47..427H} or could perhaps be from the top down with production by destruction of larger PAHs \citep{2013RvMP...85.1021T}, as shown in Fig.~\ref{fig:cgrains}.  The cold temperatures and reduced ultraviolet (UV) radiation fields ($\tau_{\rm UV} \gg 1$) in dense quiescent gas limit the destruction pathways, making in situ formation more likely, at least for smaller PAHs \citep{2021A&A...652L...9C}.   Given the widespread evidence for gas phase freezeout in this stage \citep{2007ARA&A..45..339B}, both small and large PAHs will eventually freeze on the surfaces of water-ice coated grains.
  At least 3~ppm of carbon in PAHs is implanted on top of (or mixed within) the water ice layer on dust grains \citep{2024Sci...386..810W}\footnote{\citet{2024Sci...386..810W} finds that 1\% of the interstellar carbon inventory is held in just one molecule, pyrene.}; given the likelihood of numerous hidden species, this number is likely much higher. 
These PAHs are either draping the water ice mantle, together with CO and CO$_2$, or are perhaps co-mixed.  Regardless these ices likely represent the earliest stage of the meteoritic SOM.   In \S~\ref{sec:disk} we discuss the potential for this material to contribute to the IOM.


Given the low temperatures, this phase is an important engine for implanting heavy isotope enrichment for carbon, hydrogen, and nitrogen.   In particular, significant amounts of deuterium (and nitrogen) enrichments in solids from the solar system (SOM and IOM, cometary; detailed in the Supplementary Materials) likely occur in the prestellar core of dense cold gas  \citep{2023ASPC..534.1075N}.
  TMC-1, the template site for cold PAH detections, shows evidence of pre-cursor hydrocarbons with high D enrichment  \citep[$\sim$0.1-0.3\%;][]{2022A&A...657L...5C} at a level similar to those seen in SOM and IOM (see Supplementary Materials).
  PAHs formed in this gas and condensed onto solids will be D-enriched. The current limits on the D/H ratio of benzonitrile (the simplest aromatic) are $<$1\% \citep{2025A&A...700A.281S}.

A final point is that  laboratory experiments suggest that one possible path to IOM formation is via ion irradiation (e.g. cosmic ray processing) of interstellar ices \citep{Palumbo2004, Auge2016, 2016GeCoA.189..184D}. This could potentially link to this evolutionary phase (or later phases where sources of particle radiation are present).  UV processing is less likely, as the observed hydrocarbon chemistry is more likely to lead to greater complexity in the absence of gaseous oxygen which would be released from ices in the presence of UV radiation \citep{2009ApJ...690.1497H}.

\subsubsection{Star-forming and Warm Molecular Gas}
\label{sec:starforminggas}

As noted above, many of the precursor SOM organic ices that form during the cold quiescent phases are hidden from observation.  This is because their vibrational modes are overwhelmed by more abundant molecules, as abundant organics have similar functional group stretches  \citep[e.g. CH$_3$, C-H, C-N;][]{Boogert15, Rocha2024}.
The most comprehensive inventories of hidden interstellar ices are instead found in so-called ``hot cores'' or ``hot corinos'', for high  ($>$1~M$_\odot$) and low ($\le$1~M$_\odot$) mass stars, respectively \citep{2012A&ARv..20...56C}.  These are regions of the hot protostellar envelope (or perhaps the surface of a young protostellar disk) that are heated above the sublimation temperature of water and most interstellar organic ices \citep{2022ESC.....6..597M}.  Template sources, such as Orion KL \citep[][]{2014ApJ...787..112C} or IRAS 16293 \citep{2020ARA&A..58..727J} have the most available information \citep[also Sgr B2;][]{2013A&A...559A..47B}.  For both high- and low-mass regions, one aspect stands out: for quiescent gas, the carbon inventory is dominated by saturated and unsaturated hydrocarbons and nitriles \citep{Xu_TMC_Census}.  In regions dominated by sublimated ice, organics are mainly oxygen-rich and nitriles (having a C$\equiv$N functional group).  Examples include: HCOOH, CH$_3$OCH$_3$,  CH$_3$OCHO, CH$_3$CN, C$_2$H$_5$CH, etc.  These comprise $\sim$ 3 ppm of the total carbon inventory \citep{2014FaDi..168...61B} and alongside the abundant ices (H$_2$O, CO, CO$_2$, CH$_3$OH) represent the seeds of the SOM (and, as discussed in \S~\ref{sec:disk} perhaps a portion of the IOM).   

Deuterium enrichments at the several percent level are readily observed in numerous oxygen-rich organics observed in gas with T $\sim$100~K \citep[i.e. no active fractionation;][]{2020A&A...635A..48M}.  These must originate via fractionation processes activated in the earlier and colder pre-stellar phase \citep{millar_dfrac}.  For carbon, some source-to-source variation is found with near ISM ratios for typical oxygen-containing organics, such as methyl formate \citep[][]{2014ApJS..215...25F}, while other sources exhibit $^{13}$C enrichment in the same molecule \citep{2020ARA&A..58..727J}.  In general, complex organics reflect the isotopic ratio of pre-cursor molecules (e.g., CO), with some adjustment as they grow \citep{2024ApJ...970...55I}.   We note that ISM measurements of isotopic ratios for carbon hide the observed meteoritic levels in the uncertainty (see Supplementary Materials).  

Complex hydrocarbons (PAHs) seen by radio observations in quiescent material are not readily detected in embedded protostellar envelopes \citep[][]{2009A&A...495..837G}, and there is little evidence of their direct sublimation in hot cores \citep{vanthoff2020}. However,
 AIBs, associated with PAH UV excitation followed by radiative decay are found in dense UV-illuminated regions (n $>$10$^{5}$~\cc ) known as Photodissociation Regions or PDRs of which the Orion Bar represents a template object \citep{2024A&A...685A..75C}.   
 There is some uncertainty in the overall PAH  inventory.
 The abundance of carbon locked within gaseous PAHs in the Orion Bar is $\sim$8 ppm \citep{2005pcim.book.....T}, while \citet{2023ApJ...948...55H}  require $\sim$40 ppm in PAHs for the diffuse ISM in their dust model.  Thus, the contribution of PAH to the interstellar inventory could be higher than listed in Table~\ref{tab:c_assay}.
It is possible that some PAHs are forming in regions less exposed to UV within PDRs and  
\citet{2025A&A...696A.100G} provides  evidence of bottom-up PAH formation in the Orion Bar. In addition, there is evidence of variable deuterium enrichment in the solid state carbon inventory, particularly for aliphatics \citep{2023ApJ...959...74B, Peeters2024}; although the level of enrichment is affected by the loss of H in the PDR which obscures the direct connection to formation.

\begin{textbox}[h]\section{PLANET-FORMING DISK EVOLUTION}
Rotating disks form in the protostellar phase (Class 0/I; t $\sim$ 0.5~Myr) where the energy released from accretion leads to molecular sublimation fronts that lie at greater stellar distances compared to the Class II protoplanetary disk, the phase where stellar irradiation dominates disk heating. The disk surface is irradiated by UV and X-ray radiation generated by accretion (UV, X-ray) and stellar magnetic activity (X-ray). Within the disk, high (\nhtwo\ $> 10^8$~\cc ) densities foster grain coagulation, settling, fragmentation, and drift.  Overall, grains cycle through a range of sizes, with planetesimal formation promoted in localized regions (\S~\ref{sec:planetesimalform}).  Dust evolution depletes upper layers of absorbers, and high energy photons have greater penetrating power. Over millions of years the gas dissipates.  
\end{textbox}
\subsection{Protoplanetary Disk}
\label{sec:disk}
We characterize the disk carbon inventory on two radial scales, r $\gtrsim$ 10~au (outer disk) and r $<$ 3~au (inner disk).  
For both the inner and outer disk refractory carbon is assumed to be present in nearly all models of dust continuum emission from 0.1~$\mu$m to cm wavelengths \citep[][]{2020ARA&A..58..483A}.  In these models, solid state organics represent a primary source of opacity at an assumed abundance of $\sim$ 50\% of elemental carbon  \citep{1994ApJ...421..615P, 1998A&A...332..291J, 2018ApJ...869L..45B}. This implies that the loss of condensed carbon at, e.g., the soot line (Fig.~\ref{fig:megaschematic}) could alter dust emission.  \citet{2017A&A...605L...2S} argue that the change in the millimeter-wave spectral index in an outbursting system is due to an outward gradient in the carbon content of pebbles in the inner tens of au; however, there are other solutions \citep{WangOrmel2025}.  Beyond this example, the vast majority of models assume an \textit{a priori} presence of organic-rich grains throughout the disk.  
The sidebar titled ``Planet-Forming Disk Evolution'' provides additional context.

\subsubsection{Outer disk: r $>$ 10~au}
\label{sec:outerdisk}

The abundance of disk CO is reduced by factors 5--100 below the dense ISM level provided to the disk \citep{2013ApJ...776L..38F, 2025arXiv250610738T}. This abundance reduction occurs on a timescale of $\sim 1$~Myr \citep{Zhang20, 2020ApJ...898...97B}.  Observations of disk ice absorption by JWST (in one system) by \citet{2024ApJ...975..166B} do not find excess CO (or CO$_2$, CH$_3$OH) in ices, as suggested by detailed models \citep[][and references therein]{2023ARA&A..61..287O}.  

ALMA chemical surveys reveal that the outer disk gas commonly contains elevated C/O ratios $\sim 1.5-2$ \citep[][and references therein]{Miotello19,2024ApJ...969L..21B}.  The primary tracer of this elevated carbon content is C$_2$H.  Detailed chemical models suggest that a (hidden) hydrocarbon-rich material must feed C$_2$H production \citep{Kama16a,2021ApJS..257....7B,2023ASPC..534..501M}. 
The elevated C/O ratio may be related to the "missing CO" abundance, as the carbon in CO could have been fed to hydrocarbons after CO photodestruction or could perhaps originate from UV photolysis of carbonaceous grains \citep{Bosman21}.  Both lead to the inference that the outer disk gas becomes hydrocarbon-rich over time but overall carbon-depleted, as reaction products will eventually freeze on grain surfaces \citep{Bergin16}.

The outer parts of protoplanetary disks hold conditions comparable to those of the pre-stellar core (\S~\ref{sec:quiescientgas}. Both deuterium enrichments \citep{Huang17} and carbon fractionation are present \citep{2019A&A...632L..12H,2022ApJ...932..126Y}. Models find that the isotopic fractionation of deuterium \citep{2016ApJ...819...13C} and carbon \citep{2024ApJ...965..147B, 2024ApJ...966...63Y} within the disk could have contributed to the organic precursors of the SOM/IOM. 
Given the similarity in physical conditions between dense quiescent cores and the outer disk, it is possible, although speculative, that the disk may also be a regime that facilitates simple gas phase PAH formation, with isotopic fingerprints similar to  earlier phases. These PAHs will condense onto grains and add to the primordial organic inventory.

In this light, the outer disk is potentially a prime location for the processing of primordial, and disk altered, ice mantles (water, CO, CO$_2$, oxygen-rich and nitrile organics, simple aliphatic and aromatic molecules).  
The outer surface layers of the disk are dominated by ultraviolet and X-ray radiation \citep{1997ApJ...480..344G,2003A&A...397..789V}; both of these are generally absent from dense gas during the formation of icy mantles at low temperature discussed in \S~\ref{sec:denseism}.
Although grain growth and settling to a dust-rich midplane is prevalent in disks, the process of grain growth is limited in some respects by fragmentation \citep{2024ARA&A..62..157B}.
Thus, there is a population of small ($\lesssim$ 10-100~$\mu$m) ice-coated grains that are coupled to the gas and can be lofted into UV (or X-ray) exposed surface layers, leading to both photodesorption and photodissociation of ices  \citep{ 2021ApJ...919...45B} and  potentially  to macromolecule formation \citep{2012Sci...336..452C}.
Laboratory experiments of UV exposed interstellar ice analogs suggest the initial production of SOM-like material and also IOM after subsequent irradiation \citep{1999Sci...283.1135B,  2022A&A...667A.120D, 2025A&A...702A.123K}.  There are alternatives to radiative processing of ices for the origin of the IOM; one example is  polymerization of H$_2$CO \citep[e.g.,][]{2011PNAS..10819171C}. 

In sum, disk observations, the solar system carbon gradient (\S~\ref{sec:conundrum}), and theory suggest that pebbles originating from the outer disk gas will have significant refractory organic carbon, organic ices, CO/CO$_2$ \& water ice, and silicate minerals. 

\begin{figure}
    \centering
    \includegraphics[width=1.0\linewidth]{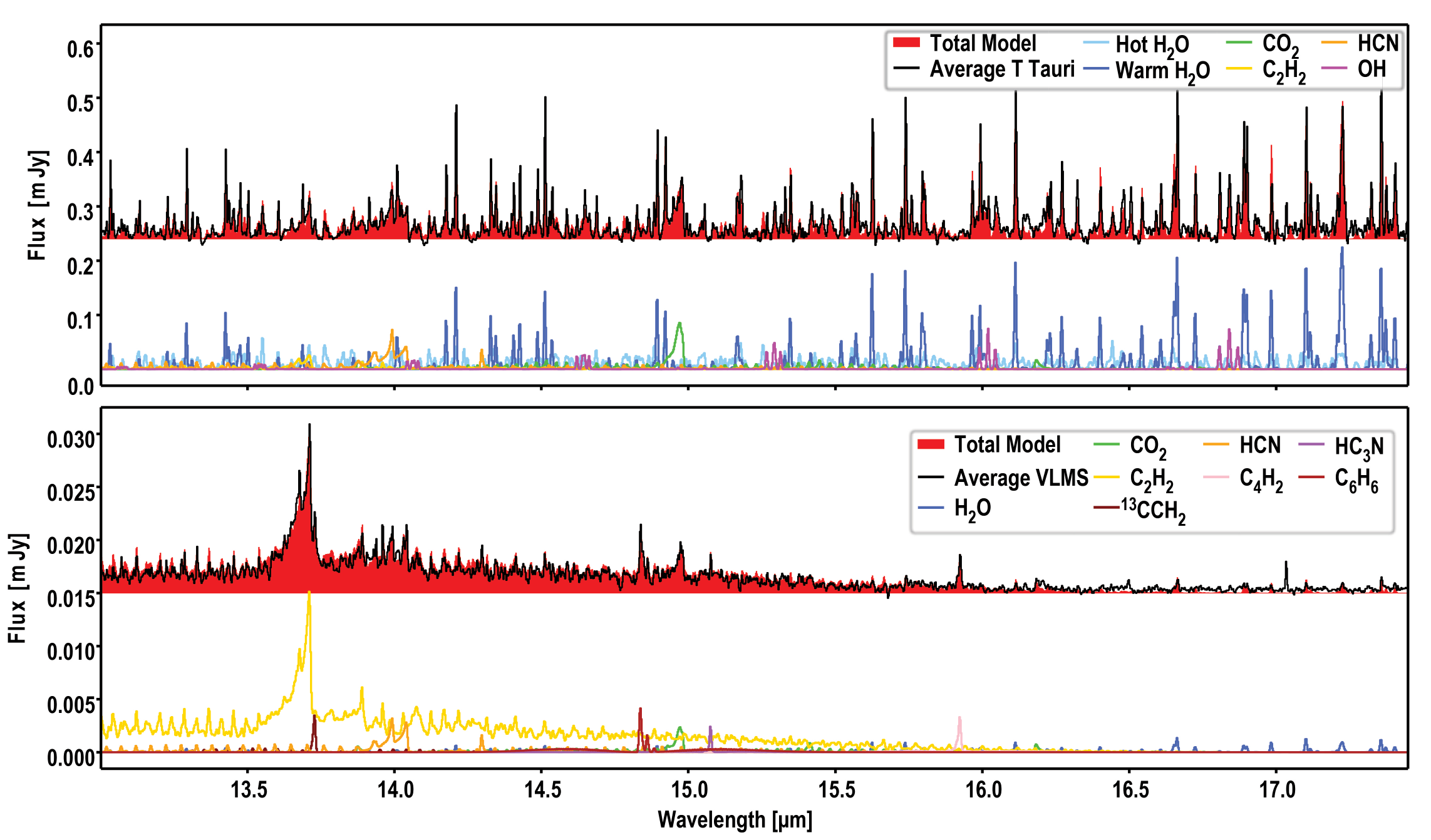}
    \caption{{\em Top:} Average JWST MIRI spectra of a sample of low mass T Tauri stars (0.2~M$_\odot <$ M$_* < 1.0$~M$_\odot$) probing the chemistry of the inner (r $<$ 3~au) protoplanetary disk. {\em Bottom:} Average spectra of a sample of Very Low Mass Stars, (M$_* < 0.2$~M$_\odot$).  In both panels the molecular emission is modeled using a slab excitation model to match the observed emission and composition.  The top spectra in each panel show the observed average spectrum and the total fit.  The bottom panel provides the speciation of the emission as a function of wavelength with composition delineated via the color coding.
Figure adapted from \citet{Grant25}. Creative Commons Attribution 4.0 License.   }
    \label{fig:jwstspec}
\end{figure}

\subsubsection{Inner disk: r $<$ 3~au}
\label{sec:innerdisk}

A striking result from Spitzer IRS surveys is that very low mass stars ($\leq$0.2~M$_\odot$; VLMS) have anomalously strong \ce{C2H2} emission relative to water  \citep{2009ApJ...696..143P, 2013ApJ...779..178P}, in comparison to disks surrounding solar mass stars.  JWST reveals that the strong \ce{C2H2} feature is a pseudo-continuum of optically thick \ce{C2H2} ro-vibrational emission, which exists alongside detections of a host of hydrocarbon emission \citep{Tabone23}.   This is shown in Fig.~\ref{fig:jwstspec} from \citet{Grant25} where the bottom panel is an average JWST spectrum of Class II VLMS disks, dominated by the \ce{C2H2} pseudo continuum and hydrocarbon/nitrile emission (i.e., C-rich); in contrast an average spectrum of a solar mass disk  is typically dominated by a host of water emission lines  (i.e., O-rich).   
 Detailed thermochemical models of systems with strong \ce{C2H2} emission are consistent with an elevated C/O ratio \citep{Najita11,Kanwar2024c2h2chem}.   In contrast,
disks surrounding solar mass stars are generally rich in water emission, which would be associated with C/O ratio $\le$ stellar; but some carbon-rich systems exist \citep{2024ApJ...977..173C}.   Dust emission at mid-infrared wavelengths is optically thick.  Thus, the molecular emission traces surface layers where T$_{\rm gas} >$ T$_{\rm dust}$.
It is uncertain whether this composition is reflective of the surface or whether it traces aspects of the midplane chemistry and bulk composition (e.g. total carbon content and elemental C/O ratio).
   Models show that the elevated C/O ratio could be consistent with the release of all the condensed refractory carbon at the soot line (Fig.~\ref{fig:megaschematic} \& \S~\ref{sec:pebbleform}).  
   However, other mechanisms activated in the gas phase  have been proposed \citep{2023A&A...677L...7M}, justifying the range given in Table~\ref{tab:c_assay}.  This will be discussed further in \S~\ref{sec:pebbleform}.

   At present isotopic ratios of key carriers in the inner disk is a field in its infancy.  JWST observations have detected $^{13}$C isotopologues of CO$_2$, HCN, and C$_2$H$_2$ \citep{Grant2023,Tabone23, 2025AJ....169..184S}.  However, in many instances, it is not clear if the emission of the main isotope is optically thick, which hides the intrinsic isotopic ratio.
   The inner disk is a potential location for the formation of IOM material. Beyond irradiation, Fischer-Tropsch  reactions (catalytic reactions on metal surfaces operating at high temperature, 500--900~K), if active, could transform abundant CO and H$_2$ into hydrocarbons \citep[][]{Kress2001}.
   However, any organics produced in the hot inner disk, via irradiation or otherwise, would have ISM D/H levels.  The fact that they do not, at least in IOM material (see Supplementary Material) favor a low temperature origin \citep[see][for greater discussion]{2011PNAS..10819171C}.

\begin{textbox}[h]\section{DEBRIS DISK EVOLUTION}
Debris Disks represent the final stage of planet formation.    Stellar radiation pressure depletes small grains requiring continuous regeneration.   High-resolution ALMA images of the thermal dust continuum often exhibit a significant substructure  attributed to the influence of hidden planets, and planets are detected in some systems.
  A key reference for this evolutionary phase is \citet{2018ARA&A..56..541H}.
\end{textbox}

\subsection{Debris Disks}
\label{sec:debris}
The carbon content of this phase (see sidebar titled ``Debris Disk Evolution'') is quite difficult to quantify.
There is residual CO vapor believed to originate from cometary outgassing \citep{2018ARA&A..56..541H}; however, this is not a major repository.  The super solar ionized and neutral carbon content in the $\beta$~Pic disk has been suggested to possibly originate from planetesimal outgassing \citep{Roberge06} and the dust emission models in some systems are attributed to carbonaceous solids \citep{2015ApJ...798...96R, 2017ApJ...840L..20L}.  In general, the presence of carbon in solid state form is inferred, but direct confirmation of the content remains elusive.

\section{CARBON IN PLANETARY BODIES}

\subsection{Solar System Bodies}
\label{sec:solarsys}

\subsubsection{small bodies: comets, asteroids, meteorites}
\label{sec:smallbodies}

\paragraph{Comets} Comets contain carbon as volatile ices and in refractory compounds. The volatile ices are primarily CO and/or CO$_2$ \citep[][]{2011ARA&A..49..471M}. The most comprehensive information available for the refractory carbon composition derives from two comet rendezvous missions, Giotto to Comet 1P/Halley and Rosetta to Comet 67P/Churyumov-Gerasimenko.  These two comets have C/Si ratios (see Fig.~\ref{fig:C_Si_solarsys}) comparable to solar and refractories are the dominant carbon carriers by factors of 2--3 \citep{Bergin15, Rubin19}; this depends on the assumed dust/ice ratio which ranges from 1-3 \citep[see][]{malamud2024uranus}.  The average composition of this material, normalized to 100 carbon atoms, is C$_{100}$H$_{80}$O$_{20}$N$_4$S$_2$ for 1P \citep{1987Natur.326..755K} and C$_{100}$H$_{104}$N$_{3.5}$ for 67P \citep{2017MNRAS.469S.506F, 2019A&A...630A..27I}, comparable to the average composition of meteoritic IOM (discussed below) and suggests a genetic link.
Refractory carbon comprises $\sim$28\% of the total mass of cometary rocks \citep{Bardyn17}.  As these values appear to reflect a more general population \citep{2021PSJ.....2...25W}, comets represent some of the most carbon-rich bodies in the Solar System.

The exact composition of the macromolecular material \citep[often called CHON particles][]{1988Natur.332..691J} is not entirely clear.  \citet{1999SSRv...90..109F} summarizes the knowledge of Comet Halley.  A portion of the refractories ($\sim$ 6~wt.\%) is composed of aromatic and long-chain aliphatic hydrocarbons.  The largest inventory is found in more macromolecular material with higher O and N content ($\sim$20~wt.\%).  Finally, there is an amorphous carbon component associated with very small grains ($< 0.05~\mu$m).

\paragraph{Asteroids and Meteorites} Spectral analyses indicate that asteroids have diverse compositional characteristics, but from the point of view of carbon, the “C” class of asteroids are the most interesting common type.  Comprising ~25-40\% of the known asteroids in the main belt \citep{choi2023taxonomic}, these low albedo objects are rich in carbonaceous material.  Recent return missions from two of these, Ryugu \citep[Hyabusa;][]{yokoyama2022samples} and Bennu \citep[Osiris-Rex;][]{lauretta2024asteroid} confirm the long-standing assumption that they are related to the most primitive  type (CI, or ``Ivuna'' type) of carbonaceous chondrites, which are the principal archive of laboratory observations on the character of extraterrestrial carbon. Based on distinct nucleosynthetic isotopic fingerprints \citep{Kruijer20}, carbonaceous chondrites and C type asteroids originated outboard of Jupiter at $>$5 au \citep{2022PJAB...98..227N}  and may include components originating from trans-Neptunian distances \citep{nguyen2023abundant}. 
 
\paragraph{Meteorites, Asteroids, and Comets in the Laboratory} Organic carbon is abundant in the matrix of carbonaceous chondrites and includes aliphatic hydrocarbons, aromatic hydrocarbons, amino acids, carboxylic acids, sulfonic acids, phosphonic acids, alcohols, aldehydes, ketones, sugars, amines, amides, and abundant insoluble kerogens \citep{sephton2002organic}. Much meteoritic organic material in meteorites is macromolecular and insoluble in typical solvents (i.e. IOM), with an average composition of C$_{100}$H$_{75-79}$O$_{11-17}$N$_{3-4}$S$_{1-3}$ \citep[][these references provide a detailed review of the carbon chemistry]{Alexander17, Glavin18}.  

The origin of complex organic molecules found in carbonaceous chondrites and C type asteroids is much debated.  As discussed in \S~\ref{sec:carbonforms} and \ref{sec:assay}  this could occur in the ISM or in the protoplanetary disk via a variety of potential mechanisms \citep{busemann2006interstellar, Alexander17}.  Modification by parent body processes, including interaction with aqueous fluids (see \S~\ref{sec:carbonlossplanet}), affects much of the material sampled in carbonaceous chondrites, Osiris-Rex/Bennu and Hyabusa/Ryugu \citep{takano2024primordial,barnes2025variety}.  Less  primitive carbonaceous chondrites and the heated interiors of asteroids are also modified by thermal metamorphism (see Supplementary Table~1), resulting in further carbon loss  \citep{sephton2003investigating,nakamura2005post,tonui2014petrographic}.

Primitive carbonaceous asteroidal material typified by Bennu (Osiris-Rex), Ryugu (Hyabusa), and CI chondrites has C/Si atomic 0.8 $\pm$0.1, 4$\pm$1 wt.\% carbon \citep{lauretta2024asteroid,nakamura2022formation}, far less than the refractory carbon concentrations evident in comets (atomic C/Si 5.5$\pm1.3$, $\sim$28 wt.\% C \citep{Bardyn17} as shown in Fig.~\ref{fig:C_Si_solarsys}.  Assuming that comets accumulate from minimally processed dust in the outer Solar System, the formation of even the most primitive carbonaceous asteroid and meteorite parent bodies must involve significant processing and carbon loss (see Supplementary Materials Table 1). 

Carbon is also an abundant constituent of certain types of cosmic dust, including micrometeorites (MMs, generally $<$2 mm and collected at Earth's surface) and interplanetary dust particles (IDPs, $<$10$\mu$m, collected with high altitude aircraft). The most extreme of these are the Ultra-Carbonaceous Antarctic Micrometeorites (UCAMMS), which have super-solar C/Si (atomic) up to 10$^2$-10$^3$ \citep{dartois2018dome} and ``chondritic porous'' IDPs with C/Si from 1--20  \citep{2005A&A...433..979M, 1995GeCoA..59.2797T}. 

Some MMs and IPDs likely derive from asteroids, but these highly carbonaceous grains are believed to be cometary dust  \citep{sandford2016organic}.  Although these particles undergo some modification during atmospheric entry \citep{chan2020organics}, they are considered to be among the most primitive and minimally processed materials available for laboratory study.
Carbon in these cometary grains comes in various forms, including poorly crystallized graphite, amorphous hydrocarbon ``globules'', and a wide variety of aromatic and aliphatic organic compounds. In many cases, refractory dust particles are enveloped by thin organic coatings (100 nm) with abundant polyaromatic and carbonyl functional groups \citep{flynn2013organic}, providing tangible evidence of their origin from irradiated ice-mantled grains.  Enrichments in D and \textsuperscript{15}N signify origin in the ISM or the coldest portions of the protoplanetary disk \citep{dartois2018dome}.  Organic species are predominantly in  disordered and nanoporous nanometer-sized polyaromatic grains \citep{dobricua2012transmission}, although less refractory aliphatic and more simple carbon compounds are also present \citep{sandford2016organic}. Overall, these cometary particles suggest less thermal maturation than equivalent carbonaceous compounds in chondrites, consistent with aqueous alteration and carbon loss on asteroid parent bodies.

\subsubsection{Terrestrial Planets}
\label{sec:Terrestrial}

The supply of carbon in terrestrial planets is essential for surface conditions, climate, and habitability, but the distribution of carbon between surfaces and interiors is also a critical aspect of interior planetary dynamics. Inventories and manifestations of carbon on the surfaces and interiors of each of the terrestrial planets, Mercury, Venus, Earth, and the Moon, are distinct, perhaps providing a hint of the even greater variety of roles carbon may play in planets across the galaxy.

Carbon in the near-surface reservoirs of the Earth is $\sim$10$^{23}$~g \citep{hirschmann2018comparative}, predominantly in sedimentary limestone. Venus has much more carbon \citep[5 $\times$ 20\textsuperscript{23}~g;][]{2013GeCoA.105..146H}
in its massive atmosphere of CO\textsubscript{2} (90 bar), created by a runaway greenhouse. Mercury also has extensive near-surface carbon, evidenced by low-reflectance terranes observed by the Messenger spacecraft, amounting to 10\textsuperscript{22}-10\textsuperscript{23}~g \citep{lark2023evidence}, a remarkable amount for a planet just 0.055 M$_{\oplus}$. Though its thin (0.6-0.9 kPa) atmosphere is predominantly CO\textsubscript{2}, Mars apparently has far-less near-surface carbon, with the atmosphere amounting to $\sim2 \times 10^{19}$~g \citep{2013GeCoA.105..146H}
and carbonates stored in sediments totaling no more than an additional order of magnitude \citep{tutolo2025carbonates}. 

The rocky interiors of the terrestrial planets in our Solar System are highly carbon depleted compared to solar or even chondritic abundances (Fig.~\ref{fig:C_Si_solarsys}), although uncertainties remain even for the Earth's silicate mantle. Due to carbon’s siderophile (iron-loving) chemistry, it is likely that the metallic cores are significant reservoirs (see the sidebar titled "Planetary Cores''). These were in large part established during the initial differentiation of the planets.  Those portions stored in silicate mantles play an important role in fostering surface conditions through volcanic exhalations of CO$_2$ and other greenhouse gases. 

The C content of the BSE remains uncertain with ranges that imply  (4.5-20 $\times 10^{23}$~g) 80\% to 95\% of the carbon in the accessible Earth \citep[that is, on the surface or potentially sampled by volcanoes;][]{hirschmann2018comparative, marty2020}.
The C content of the Martian mantle is poorly constrained. Based on assumed chondritic H/C ratios and inferred H\textsubscript{2}O contents of  Martian meteorites, \citet{filiberto2019volatiles} estimate total mantle C contents of $0.8-3 \times 10^{23}$~g, but Earth's H/C ratio is not chondritic, and Mars may not be either.  Evidence from Martian basalts that the interior is reduced suggests that the carrier phase of carbon in the Martian mantle is graphite, which limits volcanogenic venting of carbonaceous species   \citep{Gaillard22}, and also obscures direct evidence of mantle abundances. 

The carbon content of the interior of Venus is also highly uncertain. The enormous mass of CO$_2$ in the Venusian atmosphere is comparable to the total carbon in the BSE. This is consistent with early degassing during differentiation and a runaway greenhouse as near-primordial features \citep{hamano2013emergence}. However, the extraordinary young volcanic channels, the \textit{canali}, on the Venusian surface may originate from pure carbonate (carbonatite) magmas, which would signify continued CO$_2$ outgassing of the planet and could mean that the massive atmosphere is either a young feature or one that is sustained by ongoing volcanogenic outgassing \citep{trussell2025importance}.

The metallic cores are expected to be the largest reservoirs of carbon in the terrestrial planets. Seismic velocities in the cores of both Earth and Mars require considerable alloying with a light element  \citep[generally considered to be some combination of S, C, H, O, and Si;][]{Birch1952, 2021Sci...373..443S}. Based on its well-known propensity to alloy with iron, carbon is thought to contribute to the low mean atomic number of planetary cores. \citet{Li21} established a plausible upper limit of 7 wt.\% for Earth's core.  However, more recent very high-pressure experiments explore the propensity of carbon to partition into molton alloy suggest a  range of 0.1-0.2 wt.\%, or $2-4 \times 10^{24}$~g \citep{blanchard2022metal}.
 This affirms the core as the greatest terrestrial carbon reservoir, but suggests that the BE limit in Fig.~\ref{fig:C_Si_solarsys} may be overly generous. Similarly, the carbon content of the Martian core is apparently suppressed by its high sulfur content though concentrations of $\sim$0.5 wt.\% are likely \citep{tsuno2018core}, comparable to the mass in the mantle.

The graphitic deposits on Mercury's surface are thought to be remnants of an extensive  layer, formed by floatation from Mercury’s magma ocean \citep{lark2023evidence}. If the magma ocean on Mercury became graphite saturated, it is likely that its large core formed at or near graphite saturation, suggesting a potentially C-rich core amounting to 0.5-6.4 wt.\%, depending on the core's Si content \citep{vander2020constraints}. This would be comparable to that on Earth. The source of C-rich material to Mercury remains mysterious. Delivery by late accretion from the outer Solar System appears insufficient. Yet Mercury’s present position at 0.4 au would otherwise suggest that the planet accreted well inside the likely soot line for much of the disk evolution \citep[e.g., \S~\ref{sec:pebbleform};][]{Li21}, making early supply also problematic.  The high carbon inferred for bulk Mercury may point to an origin at greater heliocentric distance \citep{2015MNRAS.453.3619I}.

\subsubsection{Jovian Planets}
\label{sec:Jovian}
Relative to solar gas, the atmospheres of jovian planets are enriched in carbon and other heavy elements. For Jupiter and Saturn, direct spectroscopic observation from spacecraft (Galileo, Juno, Cassini) shows carbon enrichments of 3-4, and 9 (relative to solar), respectively \citep[][]{atreya2020deep}. For Uranus and Neptune, Earth-based spectrometry indicate enrichments may be as great as a factor of 80. Enrichments of C in the atmospheres of Jupiter and Saturn are generally considered to derive from pollution of their outer envelopes by accreted icy planetesimals or pebbles \citep[e.g.,][]{2020A&A...634A..31V}. The extreme enrichments evident in the atmospheres of Uranus and Neptune reflect their smaller mass fractions of solar gas as compared to condensed materials, making their envelopes more readily polluted by heavier elements.

 Additional carbon and other heavy elements are undoubtedly sequestered in the cores of jovian planets, but quantification has great uncertainty. Classical interior models envisioned discrete cores of condensed matter surrounded by fluid envelopes, but detailed measurements of Jupiter’s gravitational harmonics combined with increased appreciation of the mutual miscibility of metal, rock, and fluid components at extreme temperatures and pressures \citep{stixrude2025core} lead to a paradigm of more extended dilute cores \citep{2022PSJ.....3..185M}, possibly with gradational heavy element concentrations and a smaller inner core composed entirely of heavy elements \citep{stevenson2022mixing}.
 One plausible model includes $\sim$25 M$_{\oplus}$ of heavy elements in Jupiter, including no more than 3 M$_{\oplus}$ of carbon in a condensed core and 2 M$_{\oplus}$ in the molecular hydrogen envelope, with the rest in an extended dilute core \citep{2022PSJ.....3..185M}. If the accreted solid bodies contributing heavy element enrichments to jovian planets had compositions similar to comets \citep[28 wt.\% C;][]{Bardyn17}, then a feasible upper limit would be that Jupiter could contain 8 M$_{\oplus}$ carbon (Fig.~\ref{fig:C_Si_solarsys}). Despite its smaller size, the mass of the heavy elements in Saturn may be comparable to that in Jupiter \citep{helled2024fuzzy}.

 The interiors of Uranus and Neptune were formerly thought to be dominated by water ice \citep{helled2010interior}, but carbon-rich bodies are more readily reconciled with refractory carbon enrichments in small outer Solar System bodies (\S~\ref{sec:smallbodies})\citep{2024Icar..42116217M}. Reaction between organic carbon and nebular hydrogen during accretion could produce a thick methane ice mantle constituting 30-40\% of the "ice giants" mass. This scenario is also more consistent with moments of inertia than bodies dominated by rock and water ice. 

 \subsubsection{Dwarf planets/icy moons}
\label{sec:dwarfplanet}
  Carbon is an important constituent on the surfaces and in the interiors of a diverse population of intermediate-sized icy objects, including dwarf planets (Pluto, Ceres) \citep{marchi2019aqueously,tegler2010methane}, large icy moons of Jovian planets \citep{clark2010detection,combe2019nature}, and Kuiper belt objects \citep{grundy2020color}. 
 The fractions of interior carbonaceous material, $\sim$15-20 \%, can be estimated for such bodies from their moments of inertia. 
 Thermal processing of complex organics in the interiors of icy moons liberate methane, CO, and NH$_3$, which are likely sources for surface compounds on Pluto \citep{kamata2019pluto} and Titan, as well as atmospheric gases for the latter \citep{2006P&SS...54.1177A}.  The surfaces of these bodies are exposed long term cosmic ray irradiation and laboratory work suggests this may re-facilitate the production of organic matter \citep{Auge2019}.

\begin{figure}
    \centering
    \includegraphics[width=0.9\linewidth]{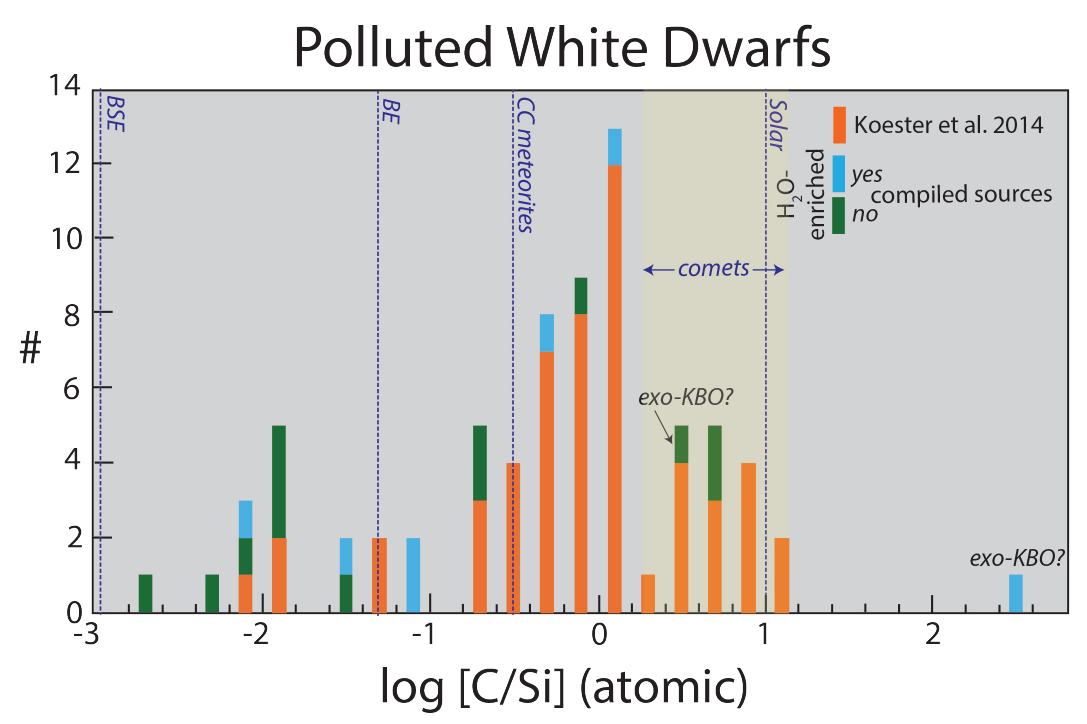}
    \caption{Histogram of [C]/[Si] (atomic) measured by optical and UV observations of the photospheres of polluted white dwarfs.  Data from \citet{koester2014atmospheric} and (“other”) \citep{gansicke2012chemical, rogers2024seven, le2024revisiting,elms2022spectral,hoskin2020white,jura2015evidence}, and from the compilation of \citet{harrison2018polluted}.  PWD that are plausibly H$_2$O-enriched are evaluated when measured O exceeds that necessary to charge balance heavy metals.  As Koester presented only C/Si ratios, data from this source are not so classified.  }
    \label{fig:PollutedWhiteDwarfsC-Si}
\end{figure}


\subsection{Polluted white dwarfs}
\label{sec:wdwarf}

 Constraints on compositions of extrasolar rocky bodies can be discerned from the compositions of photospheres of polluted white dwarfs (PWD), which reflect the consumption of debris disks (\S~\ref{sec:debris})
 sourced from satellites of those systems \citep{2016NewAR..71....9F,xu2024chemistry}. The principal sources of pollution to white dwarfs are generally thought to be planetesimals/asteroids but may extend to fragments of small terrestrial planets (cores and mantles) or even icy bodies analogous to KBOs \citep{buchan2022planets}.

Carbon/silicon ratios of planetary materials polluting white dwarfs span five orders of magnitude, from similar to the BSE up to super-solar (Fig.~\ref{fig:PollutedWhiteDwarfsC-Si}). A striking feature is that most PWD have C/Si ratios greater than carbonaceous chondrites. Although material similar to carbonaceous chondrites has been cited as a pollution source, it seems doubtful that primitive asteroids can supply the necessary mass or C enrichment because (a) the masses of an asteroid belt analogous to the main belt in our Solar System are too low to supply the necessary flux to WD over Ga \citep{Bonsor2017} and (b) thermal processing of asteroids during the red giant/AGB stage of stellar evolution should devolatilize small primitive bodies \citep{jura2010survival}.

Plausible sub-planetary sources of high C/Si include analogs of KBOs and dwarf planets. These would presumably be partly devolatilized during their journey towards their star, and this could account for observed PWD without high H$_2$O contents.  Alternatively, C-rich PWD lacking H$_2$O enrichments could reflect accretion of soot planets (\S~\ref{sec:m-r}), their fragments, or their smaller analogs.  Planets are less susceptible to late volatile loss than smaller bodies, and so the high C/Si of PWD may reflect debris disks fed in part from disruption of terrestrial planets \citep{2018MNRAS.476.3939M}.  More generally, the high C/Si ratios of PWD compared to carbonaceous chondrites indicate that small bodies greatly enriched in carbon are common in the galaxy.

\begin{figure}
    \centering
\includegraphics[width=1.0\linewidth]{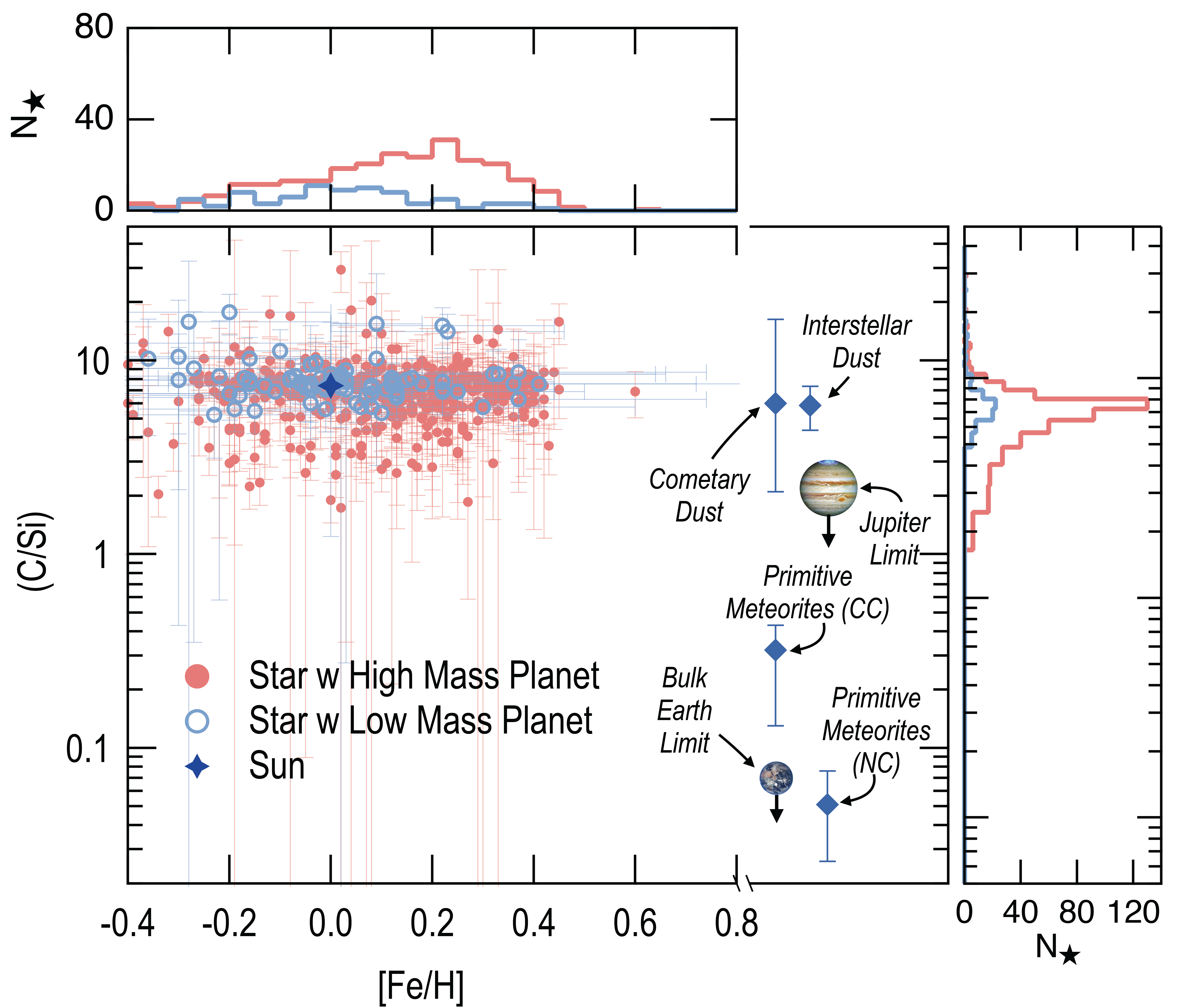}
    \caption{Stellar C/Si atomic ratio and [Fe/H] abundance normalized to solar ([Fe/H] = log(Fe/H)$_\star$ $-$ log(Fe/H)$\odot$) using the Hypatia catalog \citep{2014AJ....148...54H}.  Only systems with known planets are shown. These are broken up into whether any planet in the system is $>$30 M$_{\oplus}$ (High Mass Planet) and those systems with only planets below that mass (Low Mass Planet).  Solar system data as in Fig.~\ref{fig:C_Si_solarsys} are also shown for context.}
    \label{fig:stellar}
\end{figure}

\subsection{Stellar Composition in local neighborhood}
\label{sec:stellar}

Planetary composition is naturally influenced by the host stellar composition, and so variations in the latter are of clear interest.  For carbon an obvious question is whether there could be carbon-rich planets orbiting stars with an overall stellar C/O $>$ 1 \citep[e.g.,][]{Madhusudhan12}.  
However, the stellar C/O ratio of the majority of known planetary systems is within 50\% of the solar value of 0.5 and carbon-enriched systems are rare in the local neighborhood \citep{2012ApJ...747L..27F}. A related question is whether carbon and silicon abundances co-vary and whether the solar C/Si ratio is typical. Fig.~\ref{fig:stellar} shows C/Si ratios as a function of stellar metallicity (using Fe abundance as a proxy) for stars known to host planets within 150 pc of the Sun.  Over a range of metallicities, carbon and silicon covary, with C/Si centered on solar composition.  Systems that host higher-mass planets are correlated with higher metallicity \citep{2005ApJ...622.1102F} and a lack of low-mass planet systems at lower C/Si. The latter could be an observational bias owing to challenges in detection of lower-mass planets in the solar neighborhood. Overall, the average C/Si ratio is well above the Bulk Earth limit suggesting that there could be large, condensed carbon repositories in protoplanetary disks and exoplanets, as seen for cometary dust in our Solar System.

\subsection{Exoplanet Composition}
\label{sec:exoplanet}
\subsubsection{Mass-Radius Relation}
\label{sec:m-r}
For rocky planets and planets with modest envelopes, the relation between planetary mass (M) and radius (R) provides basic information on bulk content \citep{Seager07}. 
Interpretation of M-R plots have limitations due to compositional degeneracies, but are useful for inferring contributions from volatiles (carbon, water, H$_2$, etc.) in fluid envelopes or in planetary interiors \citep{2015ApJ...801...41R}.

\begin{figure}
    \centering
    \includegraphics[width=1.0\linewidth]{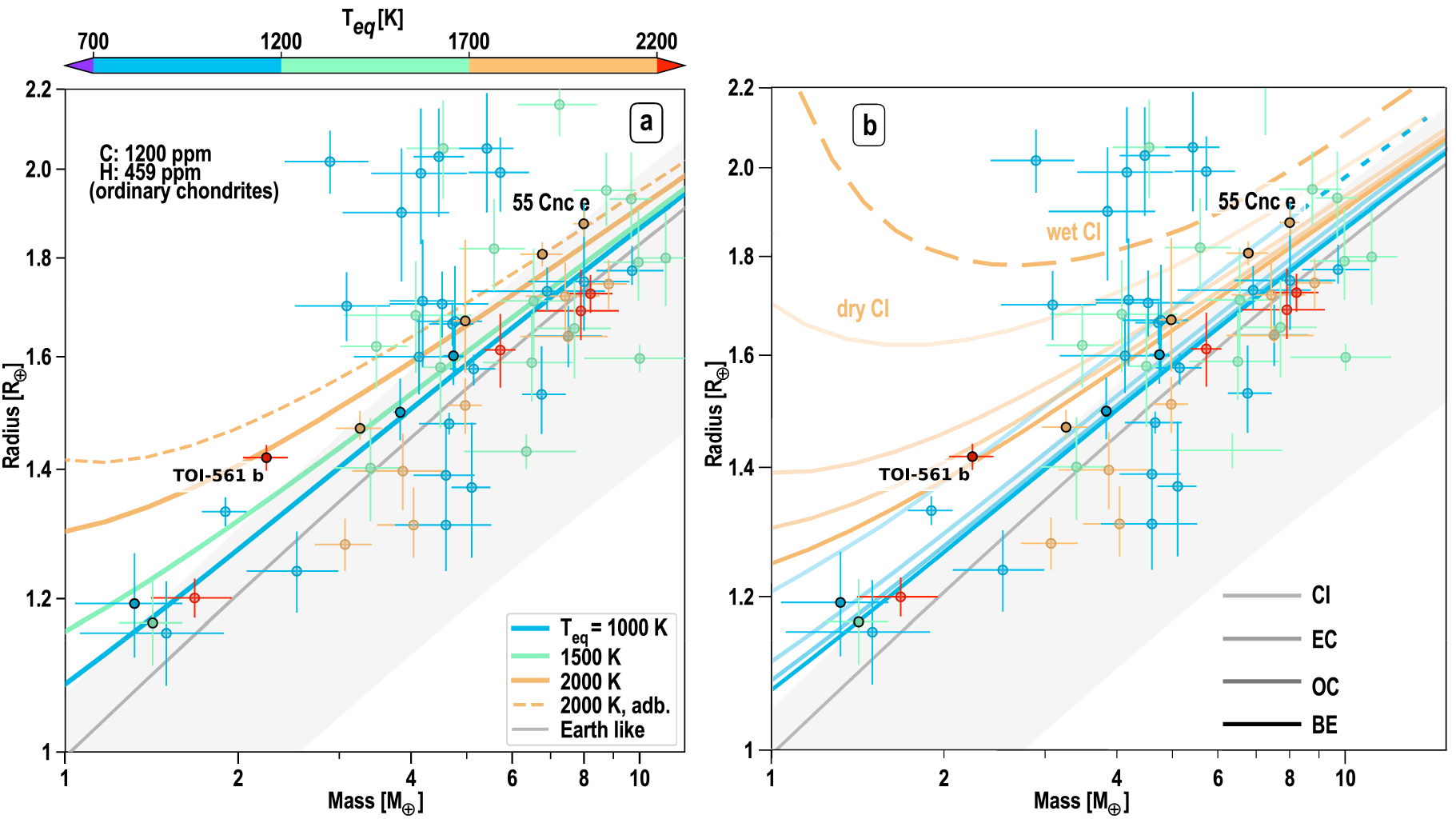}
    \caption{Mass-Radius Relations for Puffy Venuses with variable carbon content and equilibrium temperature (T$_{eq}$) from \citet{2024ApJ...976..202P}.   (a) M-R relations assuming Ordinary Chondrite level carbon and hydrogen concentrations and variable T$_{eq}$. Orange dashed line is the M-R relation for T$_{\rm eq}$ = 2000~K planets with a convective deep atmosphere to the magma ocean surface. (b) M-R relations with variable carbon content from CI (highest), Enstatite, and Ordinary Chondrite, alongside Bulk Earth (lowest) with shading values given in inset.   Contents are as follows (C/H in ppmw; parts per millions by weight) - CI:34800/0 (``dry'') or 19730 (``wet'': with more H); EC:4100/1309; OC: 1200/459; BE: 500/200) with an initial solar C/O ratio.
    The orange-dashed line is for wet-CI and the dotted line for dry-CI.  Line colors refer to T$_{\rm eq}$. 
    (a+b)  The shaded region encompasses the range an airless rocky work could occupy for  all possible Fe/Si ratios.  Planets above this region require a volatile envelope.  Exoplanet M-R with error bars are shaded by equilibrium temperature.  Planets with black circles are called out as puffy Venus candidates by \citet{2024ApJ...976..202P}. The gray line represents an M–R relation for rocky planets with an Earth-like
core-mass fraction of 0.325. Creative Commons Attribution 4.0 License.
    }
    \label{fig:m-r}
\end{figure}

Earth-like compositions are a useful starting point for M-R investigations, and departures to greater radius at a given mass are commonly inferred to signify 'water worlds' or, for sub-Neptunes, the presence of appreciable H$_2$ envelopes \citep{2022Sci...377.1211L}.  Apart from carbon-dominated worlds (see \S~\ref{sec:stellar}), the possible influence of abundant carbon on exoplanet M-R relations is relatively unexplored.
An example of the influence of significant carbon content is given in Fig.~\ref{fig:m-r}.   In this figure the gray area isolates the zone of an airless rocky world, and the planets above this zone must host a volatile component.  \citet{2024ApJ...976..202P} find that the thermochemical equilibrium for carbon-rich mantles (factors $\gg$ BSE content) leads to extended CO-rich atmospheres.  Lines for carbon-rich planets (Fig.~\ref{fig:m-r}a,b), labeled as puffy Venus's by Peng et al., are potentially consistent with a number of planetary systems.   

An important consideration is that planets born beyond the water ice line can accrete not only abundant water ice, but the full refractory carbon content of solids stable outboard of the soot line (comets: $\sim$28~wt.\%; CI:$\sim$4~wt.\%), and planets born in between the soot and water lines (Fig.~\ref{fig:megaschematic}) can be carbon/silicate rich but water poor \citep{2023ApJ...949L..17B}.
The M-R relations for these ``soot'' rich worlds are explored by \citet{2025SootWorlds} who demonstrate that this composition is consistent with many known exoplanets with a range of masses.   Although these planets are not readily distinguished from water/silicate worlds on the basis of M-R relations alone \citep{2022Sci...377.1211L}, the plausibility of C-rich worlds is informed by a greater understanding of the carbon distribution in solids in the Solar System (Fig.~\ref{fig:C_Si_solarsys}; \S~\ref{sec:solarsys}).   One limitation in improved resolution of the carbon-rich planetary structure is the lack of thermodynamic data for key hydrocarbon-rich phases at the temperature/pressures relevant for super-Earth mantles and sub-Neptune cores \citep{2024RvMG...90..259G}.

\subsubsection{Atmospheres}
\label{sec:exoplanatm}

\begin{figure}
\centering
\includegraphics[width=1.0\linewidth]{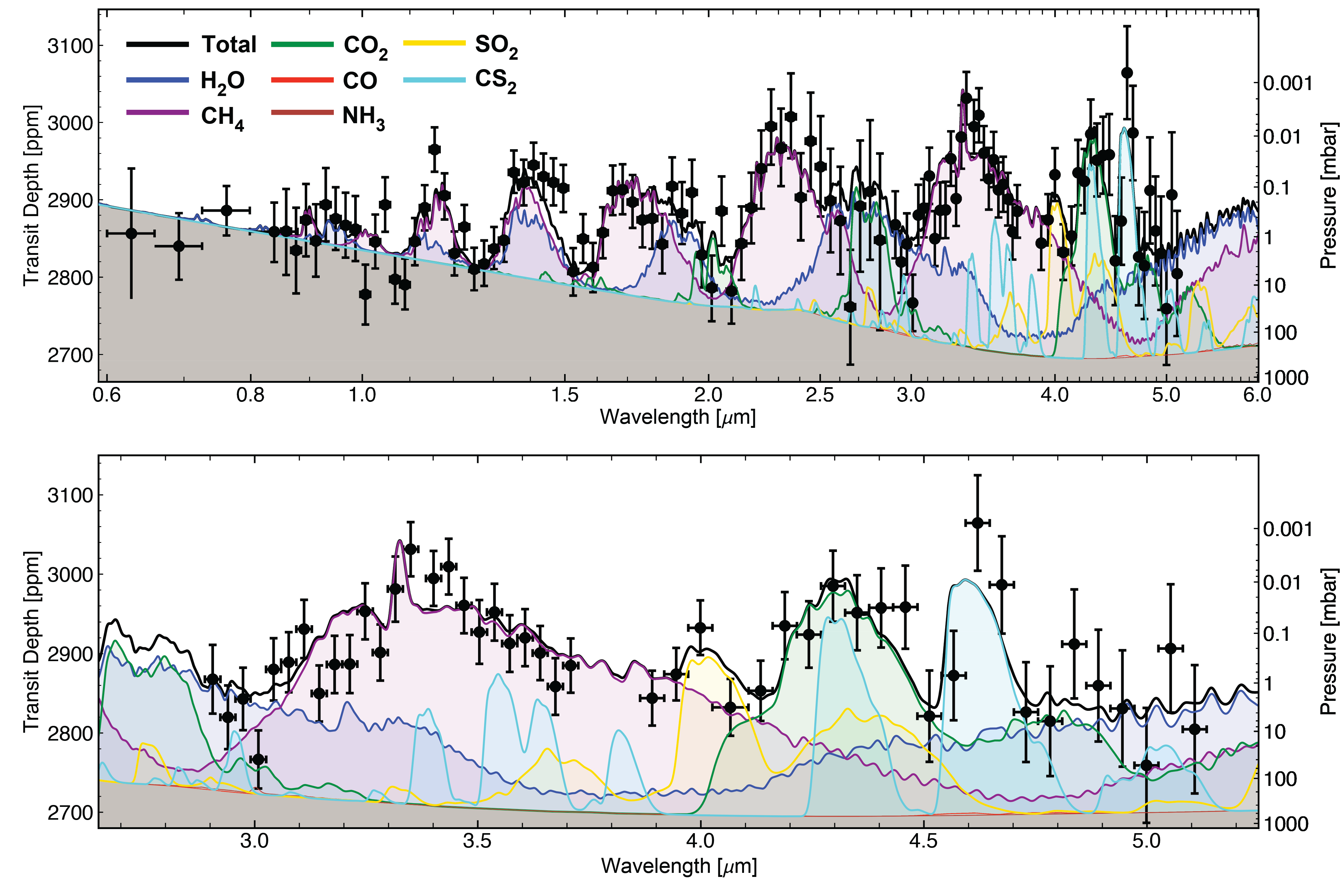} \caption{Transmission spectrum of 2.2 R$_{\oplus}$ sub-Neptune mass planet TOI-270d obtained using JWST/NIRISS and NIRSpec.  Top panel shows the transit depths at a given wavelength with best fit model shown in red.   Middle panel illustrates the contributions provided by molecular absorbers and the bottom panel is a zoom into the 3-5 $\mu$m region showing significant contributions from CH$_4$ and CO$_2$ but not CO.  Figure taken from
\citet{2024arXiv240303325B}.  Creative Commons Attribution 4.0 International License.
}
    \label{fig:toi270}
\end{figure}

The characterization of exoplanetary atmospheres is evolving rapidly with new sensitive telescopes (e.g., JWST) and techniques (high resolution spectroscopy). Here we focus on observations of carbon carriers in planetary atmospheres for planets of different mass. The central carbon carriers in planetary atmospheres are CO, CO$_2$, CH$_4$, and hydrocarbon-fueled hazes. At present, the atmospheres of Earth-sized worlds on close orbits around M stars offer the best opportunity for characterization of terrestrial world atmospheres.  Transmission observations are complicated by the presence of stellar contamination and the spectra are often non-featured, but are generally inconsistent with thick ($<$10~bar) CO and/or CO$_2$ atmospheres \citep{2023Natur.620..746Z,2025ApJ...979L...5R}, although a secondary atmosphere of CO$_2$ or CO (0.01-100 bar) has been suggested for the 1.95 R$_\oplus$ 55~Cancri~e \citep{2024Natur.630..609H}. 

For slightly higher mass planets, sub-Neptunes and terrestrial worlds with thick high-pressure atmospheres can lead to extreme greenhouse effects and therefore to a molten mantle. The magma-atmosphere exchange influences the atmospheric composition more directly than planets with a rigid surface \citep{2022PSJ.....3..127S}.  
An example of this for terrestrial planets is the suggestion of \citet{2024ApJ...976..202P} that planets with chondritic C content in the interior will have atmospheres dominated by CO, rather than CO$_2$. Hydrogen-rich atmospheres of sub-Neptune planets with carbon-rich mantles are expected to include substantial \ce{CH4} and even other hydrocarbons, which can promote the production of photochemical haze \citep{2023ApJ...949L..17B}. Hazes have been identified in the atmospheres of several sub-Neptunes \citep{Gao21}.

An intriguing result is that JWST transmission spectra of some sub-Neptune planets detect both \ce{CH4} and \ce{CO2} \citep{2023ApJ...956L..13M,2024arXiv240303325B}.  The coexistence of these is inconsistent with chemical equilibrium at the planetary temperature and has been suggested to potentially represent evidence for an underlying water ocean (Hycean world) \citep{2023ApJ...956L..13M}. However, outgassing of soot-rich mantles, as may form between the soot and water ice lines, may also produce this atmospheric signature \citep{2025arXiv250815117L}.
An example transmission spectrum of the 2.2 R$_\oplus$ sub-Neptune TOI-270-d, obtained by \citet{2024arXiv240303325B}, with detections of CO$_2$, CH$_4$, and H$_2$O, is shown in Fig.~\ref{fig:toi270}. 
This atmosphere has a mean molecular weight of $5.47_{-1.14}^{+1.25}$. 
\citet{2024arXiv240303325B} infer that TOI-270-d is a ``miscible-envelope'' sub-Neptune, with components that would otherwise be in ices in "ice giant" planets (Uranus, Neptune) instead dissolved in the atmosphere.

Finally, for larger Jovian mass planets the presence of super solar carbon abundances (with a range of C/O) appears to be common, as seen in our Solar System (\S~\ref{sec:Jovian}), although a smaller number of hot Jupiter's have sub-solar C abundances \citep{2024RvMG...90..411K}.

\section{ACCRETION, DELIVERY, DIFFERENTIATION, AND LOSS}

\label{sec:makingplanets}

\subsection{Pebbles: Formation, Drift, and Carbon Gain/Loss} 
\label{sec:pebbleform}

The evolution of solids in protoplanetary disks proceeds through the growth of sub-micron dust into larger aggregates by collisional coagulation \citep[][and references therein]{2024ARA&A..62..157B}. At early stages, van der Waals forces  promote efficient sticking, allowing grains to grow toward mm--cm sizes; in the astronomical literature these are called pebbles. However, several barriers limit further growth: the \textit{bouncing barrier}, where collisions dissipate energy without net growth; the \textit{fragmentation barrier}, where collision stresses exceed material strength; and the \textit{radial drift barrier}, whereby particles large enough to partially decouple from the gas experience rapid inward migration.   Many of these features are shown schematically in Fig.~\ref{fig:megaschematic}.  Theoretical models and laboratory experiments suggest that fragmentation thresholds are material-dependent, with icy particles expected to have higher sticking efficiencies than silicates \citep{Guttler2010}. Yet, the extent of this advantage remains debated. 
The dynamics of pebbles in gaseous disks is  highly sensitive to particle growth rates, turbulence levels, and the balance between coagulation and fragmentation.  Smaller dust can be created via fragmentation and grains could potentially cycle through a range of sizes over millions of years. Models suggest that the majority of the dust mass is found as pebbles in the disk midplane \citep{Birnstiel12,2024MNRAS.530.2731T} by 0.5 Myr or earlier.

\begin{marginnote}[]
\entry{Pyrolysis}{Process via which complex organics break down into simpler ones via the application of heat.}
\entry{Sublimation}{Pressure dependent transition of molecule from solid directly to the vapor phase.}
\entry{Photolysis or Photoablation}{Process via which molecular fragments are released from solids after exposure to high-energy UV and/or X-ray radiation. }
\entry{Oxidation}{chemical reaction of O or OH with carbon-rich grain which results in release of CO to the gas.}
\entry{Chemisputtering}{Chemical reaction of a gas-phase species with carbon-rich surface resulting in release of surface species to the gas.}
\end{marginnote}

Along the path to pebble-sized objects, carbonaceous grains are affected by multiple destruction mechanisms (see Supplementary Table~1).  The initial phase of refractory carbon loss is associated with the earliest phases of disk formation, where temperatures are high enough ($>$2000~K) to vaporize all primordial amorphous solids and, as the very young disk cools, the condensation of crystalline minerals proceeds \citep{Cassen2001, lodders03}.  In this case, the destruction of refractory organics in the hot gas would be followed by the formation of CO, rather than organics.  This initial disk then viscously expands outward, mixing with interstellar primordial grains and diluting the carbon content of solid material.   This is suggested to account for the occurrence of mineral tracers of the hot phase (e.g., calcium aluminum rich inclusions) in cometary materials returned by STARDUST \citep{2014AREPS..42..179B}.  However,   analysis of ALMA observations of disk mass and size evolution is more consistent with wind driven evolution as opposed to the viscous model \citep{Tabone2025}.  Overall, the impact of the initial hot disk, while uncertain, may be an important contributor to inner disk depletion of carbon in refractory materials.

For the macromolecular carbon carried by primordial pebbles that accrete during later cooler phases, or that drifts in from the outer disk, other destructive processes must be considered.
Chemisputtering (active only at T$>$500~K) by O and OH as water photodissociation products, and photolysis are active on UV-exposed surfaces.  To be effective, they rely on fragmentation to reduce pebble mass, as only small grains ($\lesssim 10~\mu$m) are pushed by turbulence from the midplane to UV-exposed layers.
Detailed models demonstrate even with strong mixing, chemisputtering does not  destroy carbon grains effectively \citep{Lee10, Anderson17,Klarman18}.\footnote{Chemisputtering via H atom, H$_2$, or H$_2$O is likely less effective as these are activated at even higher temperatures, while H atoms are only present in the lower density rarified disk atmosphere.}  Laboratory experiments of interstellar analog hydrogenated amorphous C exposed to UV radiation at low (10~K) temperatures find photolysis releases small hydrocarbons into the gas \citep{2014A&A...569A.119A}. This mechanism is temperature independent and thus active on the entire UV-exposed disk surface.  In regions where grains approach the fragmentation barrier, continual fragmentation replenishes small particles that are mixed into the disk surface layers via turbulence, where photolysis efficiently destroys refractory carbon. \citet{2025A&A...696A.215V} show that, when combined with processes such as sublimation, this mechanism can lead to substantial carbon depletion from pebbles near 1 au in disks around Sun-like stars.

\begin{figure}
    \centering
    \includegraphics[width=1.0\linewidth]{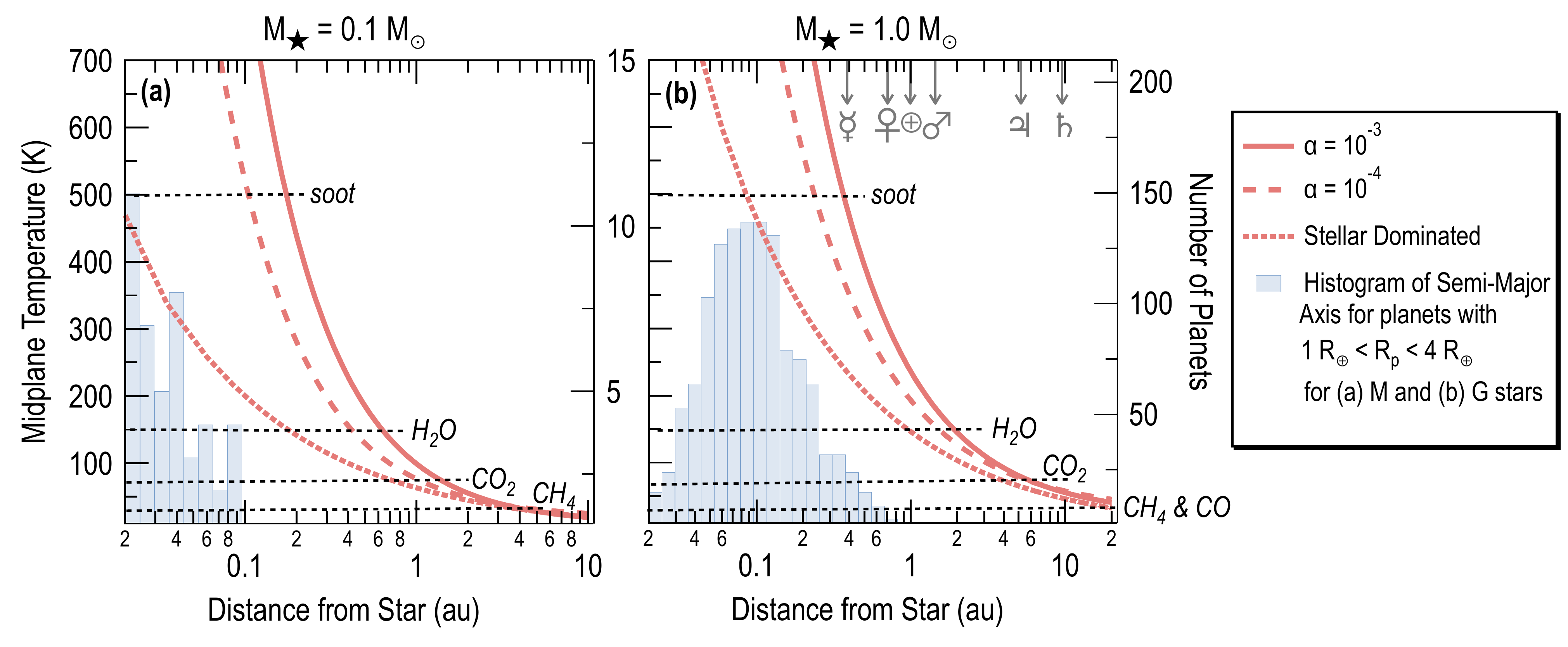}
    \caption{Plot of predicted disk midplane temperature as a function of radius for two young stars of (a) 0.1~M$_{\odot}$ and (b) 1~M$_{\odot}$. 
    Also shown are the semi-major axis distribution of planets with 1 R$_{\oplus}$ $<$ R$_p$ $<$ 4~R$_{\oplus}$ around M and G stars, respectively.
    Models of the temperature dependence with radius are taken from \citet{2023A&A...677L...7M} and are shown in red. 
 Line types show disk models heated by viscous dissipation due to accretion, with $\alpha = 10^{-3}$ 
(solid) and $\alpha = 10^{-4}$ (long dash), and a case where the temperature 
is dominated by stellar irradiation (dotted). The latter uses the same stellar parameters as Mah et al. and the dependencies from \citet{2018ApJ...869L..46D}. These three lines can also be viewed as a time sequence/evolution from younger ($\alpha = 10^{-3}$) to older (stellar irradiation dominated).
Major sublimation fronts are labeled, and the solar mass panel shows the location of planets in the Solar System using traditional symbols.}
    \label{fig:diskt-wplanets}
\end{figure}

Sublimation fronts or snowlines are important facets at influencing initial planet composition \citep{Oberg11_C_O} in part as they are active in the disk midplane and surface. In models assuming that water ice-coated pebbles fragment at higher collisional velocities, icy pebbles are relatively larger beyond the water ice line (up to a few centimeters)  than pebbles inside it (up to a few millimeters). As pebbles drift inward, they sublimate at snowlines, enriching the local gas phase with volatiles \citep{2004ApJ...614..490C}.
For carbon, the CO and CO$_2$ ice lines lie at greater distances from the star (Fig.~\ref{fig:megaschematic}), but the soot (or tar line) where the refractory carbonaceous grains are subject to pyrolysis and subsequent sublimation is found interior to the water ice line (Figs.~\ref{fig:megaschematic} \& \ref{fig:diskt-wplanets}).

\begin{marginnote}[]
\entry{$\alpha$-viscosity}{Represents  level of disk turbulence, quantifying the efficiency of angular momentum transport and controlling mass accretion rate and viscous heating.}
\end{marginnote}


The temperature of the soot line is estimated from analogs of cometary solids discussed in \S~\ref{sec:carbonforms} \& \ref{sec:smallbodies}.  These include sublimation experiments on irradiated ice/hydrocarbon mixtures
 and using lab experiments of kerogen pyrolysis and sublimation to determine degradation at temperatures $\sim$500~K
 \citep{1994ApJ...421..615P,Nakano03, 2017A&A...606A..16G, Li21, 2023MNRAS.520.2055B}. 
 Current models of interstellar dust \citep{2023ApJ...948...55H,2024A&A...684A..34Y} assume organic carbon is mixed with silicates perhaps as a carbon-rich coating.  Pyrolysis experiments of kerogens mixed with silicate minerals find differences compared to pure kerogens \citep{KARABAKAN1998}.  The likely presence of mixtures and the lack of clarity regarding the extent of the chemical processing that produces the IOM might leave some original relatively more volatile material (i.e. PAHs, hydrocarbons) intact create some uncertainty in the exact location of the soot line.
 Fig.~\ref{fig:diskt-wplanets} shows the evolving location of the ice and soot lines for a solar mass and 0.1 M$_\odot$ star. These models capture disk thermal evolution as the dominant heat source transitions from viscous accretion ($\alpha = 0.001-0.01$) to stellar irradiation~\citep{1991ApJ...380..617C,chianggoldreich97}. Based on accretion rate, $\Dot{M}$, stellar age estimates, and disk physical models this shift occurs between 0.3 and 3 Myr ($\Dot{M}$ = $10^{-7}$\; to \; $3 \times 10^{-9}$~M$_{\odot}$/yr) \citep{Hartmann16, 1999ApJ...527..893D}.  
This predicts a $\sim$500~K soot line evolving from 1.5 au to 0.1 au \citep[Fig.~\ref{fig:diskt-wplanets};][]{Hueso05}, which led to the suggestion of \citet{Li21} that the first stage of carbon loss in the Solar System occurs early. Planetesimal formation in the inner Solar System was likely initiated during the active accretion stage ($\lesssim$0.5~Myr), when the disk remained hot and both the water and carbon condensation fronts lay beyond 1 au \citep{2016Icar..267..368M,izidoroetal22}, resulting in volatile-poor, rocky planetesimals in the inner disk.


 \begin{figure}
    \centering
    \includegraphics[width=1\linewidth]{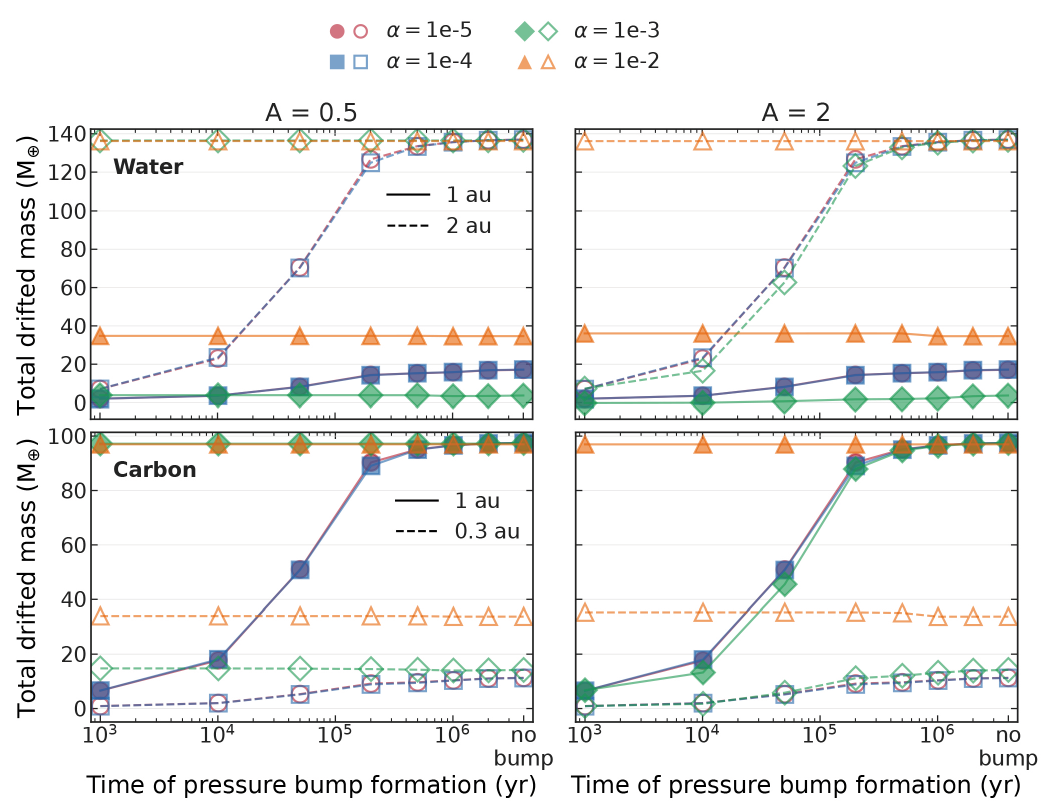}
\caption{Total mass of solids (pebbles in units of Earth mass) drifting inward as a function of the 
time of pressure--bump formation. The \textbf{top panels} shows the total mass 
crossing 2~au (solid lines and filled markers) and 1~au (dashed lines and open 
markers). The \textbf{bottom panels} shows the total mass crossing 1~au (solid 
lines and filled markers) and 0.3~au (dashed lines and open markers). These 
locations are chosen because the water snowline initially 
lies at 1.4~au and the carbon sublimation line at 0.4~au. The disk temperature profile is $T(r)\approx r^{-0.8}$ \citep{2023A&A...677L...7M}, and the surface density profile is  $\Sigma(r)\approx r^{-1}$. The initial disk mass in solids is $\sim$350$M_{\oplus}$, corresponding to an initial dust-to-gas ratio of 1.5\% \citep{2021A&A...653A.141A}. The disk cools 
with a e-fold timescale of 3~Myr, so these sublimation lines eventually 
sweep across the 1 and 0.3~au regions. In both panels, colors and marker 
shapes correspond to different values of the turbulence parameter $\alpha$ \citep{ss73}, 
while the strength of the bump is set by its amplitude $A$, defined as in  
\citet{2018ApJ...869L..46D}. The width of the pressure bump in both simulations is set to 2$H$, where $H$ is the disk scale height.  ``No bump'' points denote runs without a pressure bump. We assume that refractory carbon makes up 28\% of the solids \citep{Bergin15} and water ice 40\% \citep{lodders03}. Simulations were performed with a modified version of the Two-pop-py code~\citep{Birnstiel12}.}
\label{fig:drift}
\end{figure}

This continuous cycle of inward pebble drift, sublimation, vapor diffusion, and outward recondensation can create compositional gradients and local enhancements in surface density \citep{Stevenson88}. 
 Whereas water vapor can diffuse backwards behind the ice line and recondense \citep{Ros13}, the destruction of carbon grains is irreversible \citep{Li21} and instead molecular products (with greater volatility than water) can diffuse backward even beyond the ice line of water \citep{2025A&A...699A.227H}.  
 In Class II disks, the gas phase C/O ratio appears elevated in outer ($>$10~au) gas which has been argued to be the result of carbon grain photolysis \citep[\S~\ref{sec:disk};][]{Bosman21}.
 Carbon grain sublimation {\em may} be evident in strong \ce{C2H2} emission seen in the JWST spectra of VLMS \citep[Fig.~\ref{fig:jwstspec};][]{Tabone23, 2024Sci...384.1086A, 2024A&A...689A.231K}   and some solar mass stars \citep{2024ApJ...977..173C}. As an alternative \citet{2023A&A...677L...7M} suggest that the colder VLMS disks have closer in ice-lines and shorter viscous timescales than their solar mass counterparts, as seen in Fig.~\ref{fig:diskt-wplanets}. This can lead to a drop in C/O as water-coated pebbles arrive and an increase at later time as methane vapor, which sublimated at larger distances, advects inwards and powers hydrocarbon production.
 
One of the most striking results from the Atacama Large Millimeter Array (ALMA) is the discovery that substructures are ubiquitous features in protoplanetary disks \citep{Huang18, 2020ARA&A..58..483A, 2023ASPC..534..423B}. Largely asymmetric rings and gaps are observed in the majority of well-resolved disks, indicating that drifting pebbles are trapped at specific locations. Trapping locations are associated with enhanced pressure regions where the gas velocity becomes super-Keplerian, reducing the effects of gas drag. These bumps can arise from a variety of mechanisms -- such as the presence of planets, zonal flows, or opacity and viscosity transitions -- but in all cases they act as dust traps, halting the otherwise rapid inward drift of mm--cm pebbles \citep{2023ASPC..534..717D}.

Dust trapping in protoplanetary disks has profound consequences for both disk and planetary chemistry. By capturing pebbles, bumps can locally boost the solid surface density allowing the conditions for planetesimal formation (\S~\ref{sec:planetesimalform}). Pebble traps have also been suggested as sites for IOM formation \citep{2024NatAs...8.1257L}.
For carbon, the effect is especially important if pressure bumps form beyond the water snowline --which is consistent with several of those observed by ALMA \citep{2022MNRAS.514.6053J}. If these pebbles are trapped in outer pressure bumps, their carbon is potentially sequestered in planetesimals or planetary embryos rather than delivered inward via drift~\citep{izidoroetal22,2023ASPC..534..717D}.\footnote{Pressure bumps inside the water snowline, but beyond the soot line, would also trap carbon but ALMA observations (at present) cannot set strong constraints on structure inside $\sim$a few au.}   Fig.~\ref{fig:drift} shows that, without pressure bumps efficiently trapping pebbles, a few tens to over a hundred Earth masses of carbon- and water-bearing material drift into the inner disk ($< 3$~au) over the disk’s lifetime. This inward flux can be substantially reduced only if pebbles are trapped at early-forming ($\sim$0.1–0.5 Myr) pressure bumps, provided they avoid fragmentation and remain sufficiently large~\citep{weberetal18,stammleretal23}, or are efficiently accreted into growing planetesimals or planets.  The total mass of solids that drift across a given location depends sensitively on the relative position of the sublimation (condensation) fronts. For example, in Fig.~\ref{fig:drift}, the water snowline is initially located near 1.5~au; consequently, many more water-bearing pebbles drift across 2~au than 1~au, because icy pebbles sublimate before reaching 1~au until the snowline eventually moves inward past that radius as the disk cools. Likewise, the total mass of carbon-bearing material crossing 1~au exceeds that crossing 0.3~au, reflecting the more distant location of the carbon sublimation front during the early, hotter stages of disk evolution. Fig.~\ref{fig:drift} illustrates how difficult it is to prevent the inner disk from becoming enriched in carbon unless long-lived pressure traps operate from the earliest phases of disk evolution. Interestingly, dust rings -- potential signatures of such early traps -- have already been observed in a very young ($<$0.5~Myr) protostellar disk \citep{2020Natur.586..228S}.  


 
 \begin{figure}
    \centering
 \includegraphics[width=.75\linewidth]{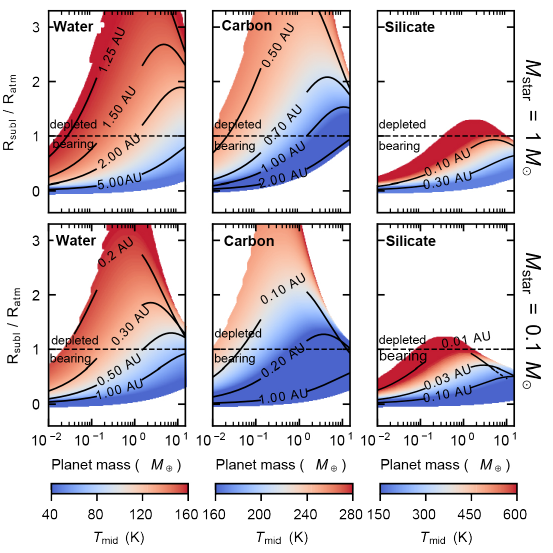}  
    \caption{This figure illustrates how the ability of a growing planetary embryo to retain or lose the solids delivered by pebble accretion depends on its mass, orbital distance, and the composition of the accreted material. It shows the ratio of the sublimation radius, \(R_{\mathrm{subl}}\) (the radial location within the planetary envelope where the local gas temperature equals the sublimation temperature of the species), to the characteristic atmospheric radius, \(R_{\mathrm{atm}}\) (the scale of the bound envelope within the Bondi sphere), as a function of planet mass. Retention of solids occurs when \(R_{\mathrm{subl}}/R_{\mathrm{atm}} < 1\), whereas loss or recycling occurs when this ratio exceeds unity.    
Results, derived following \citet{2023MNRAS.523.6186W}, are displayed  for three volatile/refractory species (top to bottom: \textbf{Water}, \textbf{Carbon}, \textbf{Silicate}) in a disk around a solar-mass star ($M_\star = 1\,M_\odot$, $L_{\rm star}=2L_\odot$; left panels) and for an M-type star ($M_\star=0.1\,M_\odot$, $L_{\rm star}=0.05L_\odot$; right panels). 
The disk structure follows a power-law model, with surface density 
$\Sigma_{\rm gas}(a) = 600\,{\rm g\,cm^{-2}}\,(a/{\rm AU})^{-1}(M_\star/M_\odot)$. The disk midplane temperature is modeled as
$T_{\rm mid}(a) = 150\,{\rm K}\,(a/{\rm AU})^{-1/2}(L_{\rm \star}/L_\odot)^{1/4}$. 
Thus, disk mass scales linearly with stellar mass, and disk temperature scales with the adopted luminosity factor $L_{\rm star}$. 
Color-coded maps show solutions from the coupled sublimation--atmosphere model; the color scale encodes $T_{\rm mid}$ (K). 
Black curves trace fixed orbital radii (labels in AU placed inline). 
The dashed line marks $R_{\mathrm{subl}}/R_{\mathrm{atm}} = 1$, separating regimes where the atmosphere is \emph{bearing} ($<1$) or \emph{depleted} ($>1$) with respect to solids at the sublimation front. 
Blank regions indicate no physical root (e.g., when $R_{\mathrm{subl}}$ lies outside the admissible envelope or $T_{\rm mid}$ exceeds the species sublimation temperature). }
\label{fig:sublimation}
\end{figure}

\subsection{Planetesimal Formation and Carbon Gain/Loss} 

\label{sec:planetesimalform}

Planetesimal formation represents a key stage in planetary growth, linking millimeter-to-centimeter dust grains to the emergence of planetary embryos -- Moon-mass objects or larger. 
The question of where and how planetesimals form remains a central uncertainty in planet formation theory. The streaming instability (SI) is widely regarded as the leading mechanism for forming planetesimals, particularly those large enough to initiate planetary growth \citep{Johansen07}, although alternatives exist \citep[e.g., vortices and zonal flows—generated by hydrodynamic or magnetohydrodynamic processes;][]{2023ASPC..534..717D}.
SI is a collective aerodynamic instability that arises from the mutual drag interaction between solid particles and gas in a protoplanetary disk. When solids drift inward due to gas drag, their relative motion with respect to the gas generates a feedback loop: local enhancements in particle density slightly accelerate the gas, which in turn reduces the headwind felt by nearby solids and allows them to drift more slowly, causing further concentration~\citep{youdingoodman05}. This feedback can lead to strong clumping of solids into dense filaments. If the local density of these clumps exceeds the Roche density, they can undergo gravitational collapse, forming $\sim$100 km sized planetesimals \citep[][and references therein]{2023ASPC..534..717D}. 

 SI is highly sensitive to local disk conditions, requiring both a sufficient abundance of marginally coupled solids and a dust-to-gas ratio above a critical threshold~\citep{Johansen07}. These requirements may be more easily met only in specific regions, such as near pressure bumps or snowlines, suggesting that planetesimal formation may be limited to discrete radial zones rather than occurring uniformly throughout the disk.  The dust continuum rings seen in the ALMA images are associated with pressure maxima \citep{Teague18b} 
capable of trapping solids (Fig.~\ref{fig:megaschematic}) and triggering SI or other instabilities.

 Planetesimals that form at different locations incorporate a range of (iron)-silicate/carbon/water contents \citep{bergin2023exoplanet, 2025SootWorlds}, as schematically illustrated in Fig.~\ref{fig:megaschematic}. 
 As planetesimals grow, they are heated by the kinetic energy of accretion and by radioactive decay of short-lived nuclides such as \textsuperscript{26}Al, leading to melting and segregation of density stratified layers, including the formation of metallic cores. For typical planetesimals 100 km in diameter \textsuperscript{26}Al heating alone is sufficient to produce melting and differentiation in the first 2 Myr after Solar System formation \citep{ghosh2006asteroidal}. 
The carbon concentrations of cores of both rocky planets and planetesimals are set during core segregation by chemical interactions between molten alloy and silicate. The ratio of mass concentration of carbon in the alloy with the coexisting silicate 
and can range from 10\textsuperscript{5} at low pressures and temperatures \citep{Armstrong14} down to near unity at very high pressures ($\sim$100 GPa) \citep{fischer2020carbon,blanchard2022metal}. It is this large preference of C for core-forming materials that has led many to conclude that metallic cores are the principal reservoir of carbon in rocky planets \citep{wood2013carbon}.  This represents a loss term for planetary surface carbon as noted in Supplementary Table 1 (under differentiation).

 As rocky bodies are heated, carbon and other volatile constituents are liberated, first by reactions producing fluids and gases and at higher temperatures, segregation into silicate melts. Hydrous fluids are produced initially at low temperatures ($\sim$100 $^\circ$C) and modest pressures, as demonstrated by aqueous alteration evident in chondritic meteorites sourced from planetesimals \citep{Lee2025}, and more extensive depletion of carbon, H$_2$O, and nitrogen occurs with increasing heating \citep[or metamorphic grade, 300-700 $^\circ$C;][]{nakamura2005post}, and these fluids are then lost through permeable networks of pores and/or fractures \citep{palguta2010fluid}.

At the beginning of silicate melting, significant amounts of volatile elements are liberated from the solid residue because of the low retention in silicate minerals. Migration of melts to the surface leads to degassing to the surface, either forming an atmosphere or allowing the volatiles to escape to space.  Some estimates of the magnitude of this effect can be found in 
iron meteorites that represent cores of fragmented planetesimals  and generally have low C contents \citep[$\sim$0.01 wt.\%; ][]{Hirschmann21}. 
 The carbon concentrations of iron meteorites do not reflect directly the compositions of planetesimal cores, but rather are products of partial crystallization of the molten cores present in the early history of the planetesimals. The compositions of these cores of planetesimals can be reconstructed from the observed low C concentrations in “magmatic” iron meteorite suites and are found to be between 0.001-0.1 wt.\%. These low C concentrations include meteorite parent bodies that based on isotopes originated from CC, or carbon-rich primitive materials. Considering the great preference of C for molten alloy compared to silicate, especially at low pressure, the low concentrations indicate that differentiated planetesimals in our Solar System lost much of the carbon originally included in primitive (chondritic; 4~wt.\%~C) solids, presumably by degassing during their accretion and differentiation \citep{Grewal2022}. 

 For planetesimals forming near or beyond the giant planets we can draw on information from the Kuiper Belt (\S~\ref{sec:dwarfplanet}) and Comets (\S~\ref{sec:smallbodies}). 
 Cold main classical Kuiper Belt objects (between 42 and 45~au; e$<$0.1; i$<$5\degree; e.g., \citet{petitetal11}) are widely regarded as the least altered remnants of the primordial planetesimal population in the outer disk. These exhibit an exponentially tapered size distribution that is expected from streaming instability–driven planetesimal formation \citep{kavelaarsetal21}.  These bodies likely record the original inventory of carbon-rich planetesimals formed in the outer disk at later stages, when $^{26}$Al heating had diminished. In contrast, comets -- especially Jupiter-family comets sourced from the scattered disk -- are predominantly kilometer-scale fragments. Their small sizes, weak mechanical strength, and volatile-rich compositions suggest that they are collisional byproducts of those larger SI-born planetesimals that either avoided substantial $^{26}$Al heating by forming late or retained volatiles in their outer layers. 

\subsection{Pebble and Planetesimal Accretion}
\label{sec:pebbleandplanetesimal}

\subsubsection{Planet Formation by Pebble Accretion}
The composition of planets growing by pebble accretion depends on the locus of accretion. As pebbles drift inward, H$_2$O ice sublimates at the water snowline and organics are volatilized at the soot line (Figs.~\ref{fig:megaschematic} \& \ref{fig:diskt-wplanets}). Embryos accreting pebbles inside these lines grow from refractory-rich, volatile-poor material; those outside can accrete appreciable volatiles.  

The fate of volatiles is also determined by the physical conditions in the accreting body’s envelope or feeding zone. Pebbles crossing a planetary embryo’s Hill sphere encounter a gaseous envelope where they may undergo thermal ablation or sublimation before reaching the core \citep[e.g.,][]{2018A&A...611A..65B, 2023MNRAS.523.6186W}.  Thus, a substantial fraction of water and/or carbon could be shed to the envelope, remaining in the atmosphere or lost back to the disk. So, the balance between direct solid delivery and envelope sublimation provides an additional filter influencing volatile delivery to growing planets (Figure \ref{fig:sublimation}). As shown in Figure~\ref{fig:sublimation}, the ratio \(R_{\mathrm{subl}}/R_{\mathrm{atm}}\) indicates whether solids delivered by pebble accretion are likely to be retained or lost. Values below unity imply that sublimation occurs within the bound envelope, allowing vapor retention, whereas values above unity result in the recycling of material back into the circumstellar disk. Overall, more massive embryos tend to have hotter envelopes and are therefore more likely to lose their accreted material, unless they form sufficiently far out in the disk.

This framework connects directly to the scenario proposed by \citet{johansenetal21}, who posited that Earth may have formed beyond the water snowline and acquired its present volatile inventory through pebble accretion. In their model, icy pebbles drifting inward are efficiently captured by growing embryos, but water and carbon are largely lost as the pebbles sublimate within the planetary envelope before reaching the core. The refractory fraction is deposited onto the planet, while the water and carbon are recycled back into the disk. This heuristic paradigm allows the formation of volatile-depleted Earth-mass planets beyond the snowline, though it could succeed only for a limited combination of disk locations and planet masses, as shown in Fig.~\ref{fig:sublimation}.

A further distinction between planetary growth modes arises between accretion during the gas-disk phase and after disk dispersal. While the disk is present, embryos can efficiently accrete pebbles, and their volatile content is subject to sublimation at snowlines and in gaseous envelopes. After dispersal, in the Debris Disk phase (\S~\ref{sec:debris}), pebble accretion effectively halts, and further growth is dominated by planetesimal and embryo collisions.

\subsubsection{The Solar System as an example of Planetesimal Accretion}
One of the most striking features of Solar System chemistry is the isotopic dichotomy between non-carbonaceous (NC) and carbonaceous (CC) meteorite reservoirs \citep{warren2011stable}. This dichotomy -- evident across multiple isotope systems, including O, Ti, Cr, Mo, Ru, and N -- is usually seen as evidence that the inner and outer Solar System were chemically isolated during the first few million years of the disk's evolution and may be a consequence of Jupiter’s early formation \citep{Kruijer20}. Separation into these two pools would inhibit transport of material, including carbon, from the outer to the inner disk \citep[see also Figure~\ref{fig:drift}]{burkhardtetal21}. However, some formation models suggest that Jupiter may not have formed fast enough to prevent large-scale mixing of materials and separation may be a consequence of a preceding pressure bump in the disk, perhaps associated with the water-snowline~\citep{Brasser20,izidoroetal22}.  

This early separation has direct implications for the low carbon inventory of the BSE compared to less-processed rocky material in the early Solar System, now represented by the chondrites (Fig.~\ref{fig:C_Si_solarsys}). We note, however that NC chondrites contain significantly more carbon than the BSE. If Earth accreted primarily from NC-type material—as suggested by its isotopic similarity to NC meteorites, then its carbon depletion must reflect additional depletion processes during and after accretion (\S~\ref{sec:planetesimalform}\ \& \ref{sec:carbonlossplanet}).

Recent Solar System formation models suggest that planetesimals did not form as a continuous radial population but in discrete rings located at pressure maxima, typically anchored to sublimation/condensation fronts such as the silicate sublimation line and the H$_2$O and CO snowlines \citep[e.g.,][]{izidoroetal22,morbidellietal25,2023ASPC..534..717D}. In this framework, inner-disk planetesimals sourced at the silicate sublimation line ($T\!\sim\!1000-1400$~K) are expected to be volatile- and carbon-poor because refractory organics are thermally destroyed and carbon resides primarily in the gas, whereas additional rings near major snowlines supply volatile-bearing material and seed giant-planet cores. An immediate implication follows for Earth’s carbon budget: if a significant share of terrestrial building blocks formed at the silicate line, a substantial fraction of BSE carbon must reflect delivery from beyond the water snowline (e.g., CC-like material or a volatile-rich impactor). FeO-rich NC material (inferred from Ordinary Chondrites and NC irons) may point to an origin of some NC planetesimals beyond the water ice line~\citep{grewaletal24}, although scenarios in which these bodies accreted from dust/pebbles that previously interacted with water (ice or vapor) are also plausible.

Together, this picture suggests that the terrestrial planets formed mainly via planetesimal accretion rather than pebble accretion~\cite[e.g.][]{izidoroetal22,morbidellietal25}. This interpretation is supported by isotopic, chronological, and dynamical constraints, which indicate that the NC–CC dichotomy was established early and persisted throughout accretion, implying that pebble drift from the outer to the inner Solar System was strongly limited \citep{Kruijer20}. Such long-term isolation, combined with thermal and chemical processing, may have naturally led to the formation of carbon-poor terrestrial planets.

\begin{textbox}[h]\section{PLANETARY CORES}
The concept of a planetary core differs by discipline and by type of planet. Earth and planetary scientists consider “cores” to be the central iron-rich metallic centers of rocky planets and differentiated planetesimals. For higher mass planets, astronomers and planetary scientists consider “cores” to be central regions of high mean atomic number condensed matter interior to low mean atomic number fluid envelopes. For these, “cores” include rocky and potentially other high mean atomic number matter and do not necessarily imply regions dominated by metallic iron.
\end{textbox}

\subsection{Carbon Loss in Planetary Bodies}
\label{sec:carbonlossplanet}
In bodies Mars-sized (0.1~M$_\oplus$) or greater, the combined heating from \textsuperscript{26}Al and accretion are sufficient to promote wholesale melting of silicate mantles, resulting in magma oceans \citep{bhatia2023role}. Magma oceans degas large fractions of the volatiles from the silicate and also promote their removal to metallic cores \citep{abe1997thermal}. The detailed apportionment depends on conditions of degassing and metal-silicate equilibration and for carbon, degassing is especially sensitive to redox \citep{Hirschmann16,Gaillard22}. Whereas under oxidized conditions, the low solubility of CO$_2$ in magma induces near complete outgassing, solubility of carbonaceous species in reduced magmas is so low that significant carbon can be retained with the silicate, potentially as graphite residue \citep{hirschmann2008ventilation,Gaillard22}.

Carbon and other volatiles outgassed to the surface of growing planetesimals and planets form atmospheres commensurate with the mass outgassed and retained, which is favored for larger bodies with lower stellar irradiance. An array of processes, including thermal and ionic mechanisms, contribute to atmospheric escape, with Jeans and hydrodynamic escape usually the most important \citep{owen2019atmospheric}.   
Atmospheres that may be gravitationally bound by their planet remain subject to ablation by impacts.  Giant impacts of \textgreater10\textsuperscript{3} km-sized  bodies near the end of the accretion process, such as that associated with formation of the Moon, can strip atmospheres from terrestrial planets \citep{genda2005enhanced}.  However, the efficiency of atmospheric removal in giant impacts depends critically on whether there was an ocean present beneath the atmosphere \citep{genda2005enhanced,lock2024atmospheric}.  During the later stages of accretion, smaller impacts of 10\textsuperscript{0}-10\textsuperscript{2} km-sized bodies are far more abundant than giant impacts and are more efficient in atmospheric removal because a greater proportion of the impacter energy translates into atmospheric acceleration \citep{Schlichting15}.   In fact, the net influence of impacts on volatile accumulations in the planet depends on the volatile content of the impactors themselves; volatile impactors such as carbonaceous chondrites can supply more carbon than they remove \citep{Schlichting15}; for such volatile-rich impactors, only smaller bodies ($\sim$1-30 km) cause effective loss for atmospheres similar to Earth. Finally, for thicker atmospheres, such as Venus, impact loss is much less efficient and has much less influence on atmosphere preservation.

\section{PLANET FORMATION: BIG PICTURE} 

\subsection{Birth in the Inner Disk}
Super-Earths and sub-Neptunes with orbital periods shorter than $\sim$100~days appear to be the most common planet arrangement around Sun-like stars, yet their formation pathways remain debated~\citep{beanetal21}. These planets span radii from $\sim$1 to 4~$R_\oplus$ and exhibit a broad range of bulk densities and potentially atmospheric properties (\S~\ref{sec:m-r} \& \ref{sec:exoplanatm}). 
An open question is whether they form locally in the inner disk (within $\sim$3~au), or originate farther out and migrate inward during the disk phase. 

In the \textit{inner disk accretion} scenario, planets assemble from solids that are locally available inside $\sim$3~au (and inside $\sim$0.1~au for lower mass stars; Fig~\ref{fig:diskt-wplanets}), including pebbles, planetesimals, and embryos. 
This does not exclude some orbital-gas driven migration during or after assembly.
Planets that assemble their cores during the early phases via pebble accretion will accrete carbon-poor pebbles, but after a $\sim$1-2~Myr the disk cools and pebbles remain carbon rich particularly for low mass stars, but carbon can then be lost if it sublimates during accretion (Fig.~\ref{fig:sublimation}).

\subsection{Migration}
\label{sec:migration}


Resonant-chain systems of super-Earths and sub-Neptunes, such as TRAPPIST-1, likely underwent  gas-driven migration and may have originated beyond the soot and snow lines~\citep[e.g.][]{shibataizidoro25}. As planetary embryos grow through planetesimal and pebble accretion and reach roughly Mars mass or larger (\(\gtrsim 0.1M_\oplus\)), type-I torques become efficient and drive rapid inward migration \citep{Ward97}. During this phase, and as they continue to grow, they likely retain much of their primordial carbon inventory --either sequestered in solids within their interiors (Fig.~\ref{fig:sublimation}) or preserved as volatiles in their accreted envelopes. Sub-Neptunes formed between the  CO$_2$ and CO  snow lines may also accrete gas envelopes with elevated carbon abundances, as oxygen is largely sequestered in ices in these regions, potentially leading to super-solar atmospheric C/O ratios \citep[e.g.,][]{2004ApJ...614..490C}. If such planets later migrate inward and experience partial or complete envelope loss (e.g., due to photoevaporation or giant impacts), their volatile-rich cores may persist and  outgas secondary atmospheres, as potentially seen in 55~Cancri~e \citep[][]{2024Natur.630..609H}.

\subsection{Influence of Giant Planets}

\label{sec:Jupiter_presence}

\subsubsection{Giant Exoplanets}

Early formation of a giant-planet core beyond the snow line should in theory intercept the inward flux of solids (Figure \ref{fig:drift}), potentially starving the inner disk and suppressing the growth of close-in super-Earths or larger \citep{lambrechtsetal19,izidoroetal22}. In this view -- plausibly relevant to the Solar System -- early formation of gas giants should inhibit the formation of large inner planets~\cite[e.g.][]{2015ApJ...800L..22I}. By contrast, later ($>$0.5~Myr; Fig.~\ref{fig:drift}) formation of gas giants, associated pressure bumps and pebble flux reduction, may not prohibit the formation of inner super-Earth or sub-Neptune planets.  Observationally, the picture is mixed: some surveys report a deficit of compact, close-in systems around cold Jupiter hosts, while others document clear coexistence \citep[e.g.][]{rosenthaletal22,2024ApJ...968L..25B,bonomoetal25}. In either case, the dynamical interplay between giant planets and low-mass planets during and after the gas-disk phase ~\citep[e.g.][]{bitschizidoro23}, and the timing of giant planet formation (\S~\ref{sec:pebbleform}), should strongly influence the carbon content of inner planetary systems.

An additional effect of cold gas giants is the implantation of planetesimals: as a giant planet grows in a gaseous disk and perturbs its surroundings, it can scatter planetesimals from the outer disk inward, injecting them across snow lines and into the terrestrial planet-forming region. Gas drag acts to stabilize these scattered bodies and facilitate their implantation in the inner system. This is the classic process for the delivery of volatiles to otherwise depleted inner regions \citep[e.g.,][]{Raymond2004Icar..168....1R,2011Natur.475..206W,Izidoro18}. In the context of exoplanets, implanted planetesimals may supplement local material during super-Earth or sub-Neptune growth, modifying their volatile budgets. In the Solar System, Jupiter likely played a dual role—both acting as a barrier to inward-drifting pebbles \citep{2014A&A...572A.107L,2015Icar..258..418M} and scattering volatile-rich planetesimals inward during epochs of growth and migration~\citep{2017Icar..297..134R}. 


In addition to implantation during the gas-rich phase, dynamical instabilities occurring after gas dispersal can also deliver carbon-rich material to the inner planetary system. In the Solar System, the giant planet instability -- often associated with the Nice model-- restructured the orbits of the outer planets and scattered a large population of icy planetesimals (CC, cometary) from beyond the giant planets inward \citep[e.g.,][]{2005Natur.435..466G,2018ARA&A..56..137N}. Overall, this instability could have delivered significant amounts of water and carbon-bearing species to the region of the terrestrial planets \citep[e.g.,][]{2023Icar..40615754J}. Some fraction of the carbon inventory in Earth's mantle and surface reservoirs may trace its origin to this epoch, though precise contributions remain debated. Unlike early gas-driven implantation, late delivery involves higher-velocity impacts and occurs in the absence of significant gas damping, making volatile retention probably less efficient and strongly dependent on impact conditions.

Isotopic evidence indicates that CC material 
contributed up to $\sim$10\% of Earth’s mass, with most estimates converging 
around 4--6\% 
\citep{burkhardtetal21}.
Assuming a bulk CC carbon content of $\sim$4~wt.\% \citep[e.g.,][]{Bergin15,Alexander17}, this corresponds to a CC-derived carbon budget of $\sim2\times10^{-4}-2\times10^{-3}~\,M_\oplus$. This is comparable to or greater than the C budget of the bulk Earth (Section~\ref{sec:Terrestrial}), but it is a maximum, as it does not account for some loss processes, such as during planetesimal differentiation (Section~\ref{sec:planetesimalform}) or impacts (Section~\ref{sec:carbonlossplanet}). 
By comparison, cometary contributions were minor: Earth’s bulk D/H ratio and heavy noble gas content imply that comets contributed a negligible fraction of the Earth's C-H-N budget
\citep{2000M&PS...35.1309M, 2013GeCoA.105..146H, Marty17}. 


\begin{figure}
    \centering
    \label{fig:evolution}
    \includegraphics[width=1\linewidth]{Fig13.png}
    \caption{Schematic outlining a strategy to detect a habitable planet by \citet{2024NatAs...8...17T}.  At top are a range of scenario’s depicted in the paper with a range of concentrations.  The bottom right provides simulated transition spectra of a temperate terrestrial planet (TRAPPIST-1f).  The spectra assume about ten JWST/NIRSpec Prism transit observations for simulated atmospheres with a range of CO$_2$ content.
    The bottom right panel On the bottom provides a simple view of a carbon cycle with liquid water working with biology to sequester carbon and depleting atmospheric CO$_2$.  Figure modified from its original version to highlight Earth's carbon cycle.  \href{https://www.nature.com/natastron/}{Reproduced with permission from Springer Nature.}
}
    \label{fig:triaud}
\end{figure}

\section{CARBON: HABITABILITY, AND EXOPLANETS}

The supply of carbon to a planet is variable, and both carbon-rich and carbon-poor bodies are expected.  The dichotomy in the composition of the inner disks between very low mass stars (VLMS, carbon-rich) and solar mass stars (predominantly water-rich) hint at the presence of hidden carbon-rich inventories (Fig.~\ref{fig:jwstspec}, \S~\ref{sec:disk}). At face value if this dichotomy relates to the destruction of refractory organics in the midplane then perhaps planets born in low mass stellar systems will be carbon-poor and solar mass systems water-rich.  However, as discussed in \S~\ref{sec:pebbleform},
alternate scenarios that feed carbon as ices from the outer disk are proposed, which would lead to refractory organic-rich planets.  The simplest statements that can be made at present is that if planets are born beyond the water iceline, then they will accrete large amounts of refractory carbon alongside water \citep{2025SootWorlds}, modulo the potential loss during accretion onto the planet (\S~\ref{sec:pebbleandplanetesimal}) and that the presence of a pebble trap to regulate carbon supply  is likely a central aspect in the formation of silicate-rich worlds.

The potential for planetary habitability is typically considered in terms of orbital location within the stellar habitable zone \citep{kasting1993habitable}, but as emphasized by \citet{Bergin15}, a prerequisite for a planet sustaining a habitable climate and surface environment is having a sufficient supply of life-essential elements. In the case of carbon, too little would not support development of prebiotic chemistry or to supply needed greenhouse CO$_2$, whereas too much could lead to a runaway greenhouse, as in Venus, because the CO$_2$ volcanogenically outgassed to the atmosphere would exhaust the supply of fresh silicate rocks, also provided by volcanic activity, to remove it.  For an Earth-like planet around a Sun-like star near 1 AU, \citet{foley2018carbon} find that planetary carbon budget, exclusive of the core, is between 10\textsuperscript{21}-10\textsuperscript{23} grams, or 0.01-1 times the amount present in today's accessible Earth. However, if a planet has plate tectonics, which supplies additional fresh rock to the surface by formation of mountains (and thereby promotes drawdown of atmospheric CO$_2$ by weathering to form carbonate minerals), inventories up to 10\textsuperscript{25} grams (100 $\times$ present Earth abundance) could potentially avoid runaway greenhouse conditions.  Even such great carbon budgets are small compared to cometary or chondritic abundances, highlighting the importance of carbon loss to establishment of habitability on terrestrial planets.

Early Mars was apparently habitable, despite the relatively small fluxes of volcanogenic CO$_2$ \citep{hirschmann2008ventilation}.  The recent detection by the Curiosity rover of carbonates in Martian soil \citep{tutolo2025carbonates} indicate a potentially thicker CO$_2$ greenhouse on young Mars, but it was likely not sufficient to maintain habitability without contributions from other gases, such as H$_2$ or sulfur species.  

Understanding carbon cycles and habitability for more exotic planets is challenging, as are related questions of detection and characterization. A recent suggestion is the somewhat paradoxical criterion that low atmospheric CO$_2$ could be a signature of habitability (Fig.~\ref{fig:triaud}). An active carbon cycle, including dissolution in a surface ocean and weathering of fresh silicate rocks, draws down CO$_2$ mixing ratios \citep{2024NatAs...8...17T}. This assumes that the main volcanogenic gases are H$_2$O and CO$_2$, which would not apply to planets with reduced interiors \citep{Gaillard22}. For super-Earths,  sub-Neptunes, and planets that accreted significant organics, another potential path to habitability could be climate modulation by formation of hydrocarbon hazes \citep{Arney16,2023ApJ...949L..17B}.

In addition to the beneficial effects of a pebble trap in the early Solar System regulating carbon influx,  one possibility is that Earth’s habitability was favored by its long assembly \citep{Chambers13}. Earth completed formation between 30 and 150~Myr, after the Solar System onset~\citep{Nimmo15}. Repeated embryo-embryo impacts over tens of Myr probably drove multiple cycles of magma-ocean formation, degassing, and atmospheric stripping, promoting carbon loss. By contrast, close-in super-Earths, particularly those in resonant chains, likely assembled much faster during the gas disk phase~\citep[e.g.][]{Izidoro17}. Shorter growth times and stronger gravity from the beginning would diminish the cumulative efficiency of escape- and impact-driven carbon loss and favor retention of volatiles. This distinction could yield a first-order difference between Earth and typical super-Earths in carbon depletion, with corresponding implications for surface conditions, including climate and atmospheric chemistry potentially conducive to prebiotic chemistry.

\begin{summary}[SUMMARY POINTS]
\begin{enumerate}
\item Carbon evolution in solar systems begins with carbonaceous grain formation in AGB stars and in the dense ISM followed by substantive carbon loss in small dust and pebbles in the disk, within planetesimals, and in planets (\S~\ref{sec:assay} \& \ref{sec:makingplanets}).
\item 
Observations from the Solar System elucidate key processes of carbon distribution and evolution.  The Solar System exhibits a strong C/Si gradient, from carbon-poor rocky bodies in the inner parts to cometary belts where bodies hold up to 28\% of their refractory mass in carbon.  Much of this gradient is governed by irreversible sublimation/photolysis of refractory solids at the soot line, which migrates through the inner disk with time (\S~\ref{sec:conundrum}, \ref{sec:solarsys} \& \ref{sec:pebbleform}).
\item The high C/Si ratios, above that seen in carbonaceous chondrites, inferred for materials accreted onto the surfaces of polluted white dwarfs indicates that small bodies greatly enriched in carbon are common in the galaxy.
\item The delivery of carbon into the inner disk via pebble drift is pervasive and, unmitigated, should deliver abundant carbon. Drift can be stopped by the formation of a strong pressure bump inside $\sim$5~au (for a solar mass star) that forms in $\lesssim$0.5~Myr and is sustained throughout the major drift phase.  This must have been the case in the Solar System, but may not be common in the galaxy (\S~\ref{sec:pebbleform} \& \ref{sec:Jupiter_presence}).  
\item The fate of carbon-rich pebbles accreting to a planet depends on the temperature profile of the envelope: planets embedded in cooler disks retain accreted refractory carbon, whereas carbon delivery is limited in warmer parts of the disk.  For planetesimals, the retention of carbon depends on the time of accretion due to the temporal decay in $^{26}$Al heating (\S~\ref{sec:planetesimalform} \& \ref{sec:pebbleandplanetesimal}).
\item Resonant planet chains theorized to form beyond the water ice line, or perhaps the soot line, are more likely to preserve the volatile inventory accreted at birth, even in the case of hydrogen loss.  These systems could be key to understanding the relationships between formation and composition (\S~\ref{sec:migration}).
\item The carbon architecture of our Solar System is strongly linked to the formation of Jupiter.  Other solar systems, lacking close-in giant planets, may be quite different, and very carbon-rich rocky worlds and cores of sub-Neptunes may be common, as supported by recent observations (\S~\ref{sec:exoplanet} \& \ref{sec:Jupiter_presence}).  
\end{enumerate}
\end{summary}

\begin{issues}[FUTURE ISSUES]
\begin{enumerate}
\item 
The presence of identified individual PAH molecules in the dense cold ISM is now established; understanding the mechanisms of formation, the chemical variability, and potential links to meteoritic/cometary organics represents a central astrochemical and cosmochemical frontier.
\item Many young protoplanetary disks have super-solar C/O ratios in the outer and inner regions. Observations and novel models that document and explain the sources of excess carbon and/or mechanisms of oxygen depletion will illuminate hidden inventories of carbonaceous solids in the early stages of planet formation. 
\item The success of JWST in revealing the composition of sub-Neptune-sized planets is an important milestone.   Looking ahead, telescopes on the ground (the Extremely Large Telescope; ELT) and future observatories (e.g. Habitable Worlds Observatory) coupled with an improved understanding of interior structure will provide a better understanding of the distribution of carbon in diverse planets. 
\item  In the Solar System, the mission to the Trojans and, hopefully, a future mission to Uranus will elucidate important aspects of carbon supply and distribution.
\item Understanding the habitability of carbon-rich worlds is crucial to capture the diversity of potential outcomes.
\end{enumerate}
\end{issues}

\bigskip
\section*{DISCLOSURE STATEMENT}
The authors are not aware of any affiliations, memberships, funding, or financial holdings that
might be perceived as affecting the objectivity of this review. 

\section*{ACKNOWLEDGMENTS}
MMH and EAB are grateful to Geoffrey Blake, Fred Ciesla, Jackie Li, and Eliza Kempton for a decade of educational discussions and a true interdisciplinary collaboration. This manuscript would not exist without their inspiration. EAB is also grateful for discussions with M. Colmenares, J. Kanwar, the JDISCS team for placing JWST results in context, and the GOTHAM team for helping to move Astrochemistry in new directions.  EAB is grateful to Alessandra Candian for insightful discussion on grain composition/structure and for providing a great figure.  We also acknowledge B. Benneke, B. Peng, and A. Triaud for allowing use of original figures.  
EAB acknowledges support from NASA grant 80NSSC24K0149, 80NSSC20K0259, and NSF Award No. 2509735. MMH acknowledges support from NSF grant EAR2317026.  A.I. acknowledges support from NASA grant 80NSSC23K0868.

%
\bibliography{z}

\newcommand{\beginsupplement}{%
  \setcounter{table}{0}
  \renewcommand{\thetable}{S\arabic{table}}%
  \setcounter{figure}{0}
  \renewcommand{\thefigure}{S\arabic{figure}}%
}
\appendix
\beginsupplement
\section*{Supplementary Materials}
\section{Evidence for Cold Origin of Primordial Materials}

\subsection{Isotope Ratio Measurements in Primary Carbon Carriers}

Soluble organic matter (SOM) and insoluble organic matter (IOM) in meteorites alongside cometary refractory organics have measurements of heavy isotopic enrichment for C (\ce{^{13}C}), H (D), and N (\ce{^{15}N}) that provide strong evidence of an origin in a cold medium \citep[see discussion in][]{2011PNAS..10819171C}.   This is because the chemistry of the ion-molecule reactions that lead to enrichment (or depletion) of the heavy isotope requires T $<$~70 ~ K to operate \citep{millar_dfrac, Langer1984, 2015A&A...576A..99R}.  Since, the formation of carbonaceous solids via combustion within the envelopes of AGB stars is found in much warmer gas \citep{1989ApJ...341..372F, 1992ApJ...401..269C} this points to an origin 
 either in the dense pre-stellar cloud or in the outer regions of a protoplantary disk (both are presented in \S~2 of the main manuscript).  These two phases (or locations within the disk) are dense (i.e., short reaction timescales) and cold with temperatures below 70~K.   Below, we provide the measurements within each component and in sidebar entitled ``Isotopic Fractionation'' list motivating and background details.  

 \begin{textbox}[h]\section{ISOTOPIC FRACTIONATION}

Isotopic fractionation is a key process linking the origin of solar system solids and the chemical processes active in earlier stages.  It results in heavy isotope enrichments for hydrogen, carbon, and nitrogen in organics  \citep{Ceccarelli14, 2023ASPC..534.1075N}.  Chemically this derives from a disparity in the zero point energy between lighter and heavier forms (e.g. H$_2$ and HD; $^{12}$CO and $^{13}$CO).  Isotope fractionation is effective below the temperature (generally $<$~30--70 K ) where differences between zero point energies are important
\citep[][]{2015A&A...576A..99R}. Due to the cold  ($<$ 70~K) temperatures, chemistry in the dense ISM and the outer parts of protoplanetary disks produces heavy isotope enrichments in water and first generation organics [e.g., (D/H)$_{\rm H_{2}O, organics}$ $>$ (D/H)$_{\rm H}$]; additional fractionation may also be linked to isotopic selective self-shielding which requires the presence of layers with T $\lesssim$ 100~K to be implanted in ices \citep{2023ASPC..534.1075N}. Additional information can be found in Aikawa et al., this volume. 
\end{textbox}

Before summarizing, we note that for the ISM  isotopic ratios are reported in the absolute sense, e.g. $x(^{12}$CO)/x($^{13}$CO) or $x$(HDO)/$x$(H$_2$O), where $x$ denotes abundance, with accuracies of $\sim$10-20\%.  Meteoric analyses have significantly lower uncertainty and report ratios in {\em per mille ($\permil$)} relative to the reference standard with errors at 1-2$\permil$.  Thus $\delta(^{13}$C) $\permil$ = [($^{13}$C/$^{12}$C$_{sample}$) - $^{13}$C/$^{12}$C$_{REF}$)]/($^{13}$C/$^{12}$C$_{REF}$) $\times 1000$.
Reference (REF) values are as follows.  For Earth Vienna Standard Mean Ocean Water (VSMOW) is (D/H)$_{\rm VSMOW}$ = 1.5575$\pm$0.0008 $\times 10^{-4}$ and the ISM (Hydrogen) is D/H = $2.3\pm0.2 \times 10^{-5}$ \citep{2006ApJ...647.1106L}; the difference between VSMOW and ISM D/H implies that Earth's water is D-enriched with a cold origin \citep{1981A&A....93..189G, millar_dfrac}.   For carbon, the Earth standard (Pee Dee Belemnite, PDB) is $^{12}$C/$^{13}$C is 89.9$\pm$0.3 \citep{2021NatPh..17..889F} and the local ISM standard is 76.2$\pm$1.9 \citep[][]{Stahl2008}, higher than the value of $\sim$70 commonly adopted for nearby molecular clouds \citep{2005ApJ...634.1126M}. The difference between ISM and the solar system is due to 4.6 Byr of galactic chemical evolution.
In our discussion, we focus our discussion of isotopes on hydrogen (largest signature) and carbon (the focus of this review). However, nitrogen isotopic ratios are also important links to earlier phases \citep{2010GeCoA..74..340M, 2019A&A...632L..12H}, but for the sake of brevity we only discuss hydrogen and carbon fractionation.

\begin{extract}
{\bf Carbon-Oxygen ices:} Measurements of  CO and CO$_2$ isotopes in ISM ice and cometary volatiles (67P) show no evidence for significant carbon fractionation \citep{2023NatAs...7..431M, Hassig2017}.   Although measurements of carbon isotopic ratios of oxygen rich organics sublimated in hot cores show evidence for $^{13}$C enrichment \citep{2022ApJ...939...84B}.  These organics (e.g., CH$_3$OH) are believed to form from CO and their origin is uncertain at present.

{\bf Soluble Organic Matter (SOM):} The SOM carries isotopic enrichments of deuterium (comparable to the factor of $\sim$10 D enrichment of Earth's water) and nitrogen that link its formation to simpler water and organic ices that initially form in the dense ISM \citep[][]{2012E&PSL.313...56M, 2017OLEB...47..249P}.  For carbon the SOM organics exhibit small carbon isotope fractionation levels for different organics \citep{Glavin18}.
For instance, amino acids are overall enriched in $^{13}$C \citep[$\delta(^{13}$C) $\sim$ 0-30$\permil$][]{2004GeCoA..68.4963P}, but there are differences with the functional groups (for alanine - one amino acid) with the amines (nitrogen single bonded to one or more carbon atoms; $\delta(^{13}$C) $\sim$ 140$\permil$) enriched relative to the Earth standard  and the methyl group (carbon bonded to three H atoms)  depleted ($\delta(^{13}$C) $\sim$ -30-0$\permil$); these are linked to ISM fractionation \citep{Chimiak2021}.  

 {\bf  Polycyclic Aromatic Hydrocarbons (PAHs):} PAHs are estimated to have deuterium enrichments that may be correlated with formation in cold material \citep{Peeters2024}.  However, the rate of H loss exceeds that of D due to UV photoextraction; elevated  D/H ratios (relative to hydrogen in the ISM) may be dominated by progressive hydrogen loss
\citep{1989ApJS...71..733A}.  \citet{2023Sci...382.1411Z} find that the carbon isotope ratios of specific PAH molecules in samples from Ryugu (Hyabusa) and the Murchison meteorite and argue that these reflect formation in the dense cold (T $<$ 30~K) interstellar medium.

{\bf Insoluble Organic Matter (IOM):}  IOM is associated with D-enrichments \citep[D/H $\sim$ 1--7 $\times$ 10$^{-4}$;][]{busemann2006interstellar, Alexander10} that hint at formation in earlier colder phases in the dense ISM or outer regions of the protoplanetary disk.  Bulk meteoritic IOM  and 1-2 ring aromatics are depleted in $^{13}$C \citep[$\delta(^{13}$C) $\sim$ -25$\permil$;][]{Alexander1998, Sephton03}.  Cometary grains
sampled by the STARDUST mission hold similar $\delta(^{13}$C) values to bulk IOM \citep{2006Sci...314.1724M}.
\citet{lawrence2025carbon} find a correlation between C and O isotopes in different classes of carbonaceous chondrites that argues for modifications and perhaps genesis of IOM in different reservoirs in the solar nebula.
\end{extract}

\subsection{Primary Carbon Fractionation Pathways in Cold Gas}
Since this review has a primary focus on carbon we outline here the primary fractionation reactions.   The central pathway is through $^{13}$C$^+$ $+$ $^{12}$CO $\rightarrow$ $^{12}$C$^+$ $+$ $^{13}$CO + 35~K which places $^{13}$C into $^{13}$CO \citep[by a factor of 2-3, $\delta(^{13}$C) = 500-660~$\permil$;][]{Langer1984} 
  and leaves a portion of the hydrocarbon pool \citep[e.g. C$_2$H, C$_3$H$_2$;][]{Loison2020} forming from $^{13}$C depleted gas (i.e. $^{12}$C/$^{13}$C$_{\rm hydrocarbon}$ $\gg$ $^{12}$C/$^{13}$C$_{\rm ISM or PDB}$, $\delta(^{13}$C) $\ll$ 0).  However, there is an additional fractionation reaction linked to C$_3$, which forms from atomic carbon ($^{13}$C + C$_3$ $\rightarrow$
$^{12}$C + $^{13}$CC$_2$ + 27~K).  C$_3$ is a pre-cursor molecule to numerous gas phase hydrocarbons, and this reaction leads to a $^{13}$C enrichment (by a factor of $\sim$2 or $\delta(^{13}$C) $\sim$500~$\permil$) in numerous gas phase hydrocarbons \citep{Colzi2020, Loison2020}.
These disparate pathways, and growth to larger sizes (such as naphthalene, C$_{10}$H$_8$), likely add together to dilute the carbon isotopic fingerprint of this phase  to the meteoritic level  and might account for the findings of different isotopic enrichments in amino acid functional groups \citep{Chimiak2021} and additional signatures in PAHs from Ryugu (Hyabusa) samples \citep{2023Sci...382.1411Z, Yabuta2023}. 

\section{Condensed Carbon Gain and Loss}

Supplementary Table 1  provides a summary of the key mechanisms active throughout the process and the size scale of the solids on which they operate.   The size scale where various processes are relevant relate to two aspects: (1) whether the solids are heated by the star/accretion ($\le$ cm)  or by internal heating  ($^{26}$Al; $\sim$km-scale) and (2) whether grains are dynamically coupled to gas motions which may elevate solids to surface layers where photons are present. Overall, each of these processes listed has some relevance, albeit at different evolutionary stages.

\setcounter{table}{0}
\renewcommand{\tablename}{Supplementary Table}
\begin{table}[h]
\tabcolsep7.5pt
\caption{Condensed Carbon Evolution: Gain and Loss Terms}
\label{tab:gain-loss}
\begin{center}
\begin{tabular}{|l|r|c|c|c|c|c|}
\hline
\multicolumn{1}{|c}{Process$^{\rm a}$} & 
\multicolumn{1}{|c}{C Carrier} & 
\multicolumn{1}{|c}{Size } & 
\multicolumn{1}{|c}{T (K)} & 
\multicolumn{1}{|c}{Location$^b$} & 
\multicolumn{1}{|c}{Phase$^{\rm c}$} & 
\multicolumn{1}{|c|}{Ref.$^{\rm d}$}\\ 
\multicolumn{1}{|l}{} &
\multicolumn{1}{|c}{} &
\multicolumn{1}{|c}{} &
\multicolumn{1}{|c}{} &
\multicolumn{1}{|c}{} &
\multicolumn{1}{|c}{} &
\multicolumn{1}{|c|}{} \\\hline
$+$combustion & PAH, aliphatic & $\le$0.1 $\mu$m &  $\sim 10^3$ & stellar env. & AGB$\star$ & (1)\\
$+$freeze-out & O-rich org. & $\le$0.1 $\mu$m & $<$ 30 & deep surface & dc, ppd &  (2) \\
 & CO, CO$_2$ ice &  &  &  &  &\\
 $+$freeze-out & small PAHs & $<$10 $\mu$m & $<$ 100 & deep surface & dc,ppd? & (3) \\
 $+$photolysis & org. $\rightarrow$ refr. org.& $<$10 $\mu$m & 30--300 & surface & ppd & (4) \\
 $-$ablation & refr. org. & $\le$0.1 $\mu$m & $>1000$ & ISM & shocks & (5)\\
$-$sublimation & O-rich org.,  & $\lesssim$1 cm &  30--120 & midplane &      ppd   &   (6)\\
 & CO, CO$_2$ ice  &  &   &  and surface &     &   \\
 $-$sublimation/ & refr. org.  & $\lesssim$1 m$^e$ &  $\gtrsim$ 500 & midplane + &      ppd   & (7)\\
 \hspace{6.5pt}pyrolysis & & &  & surface &       &\\
  &  & $>$1000 km &  $\gtrsim$ 500 & plan-emb &      ppd   & (8)\\
 $-$photoablate & refr. org.  & $\lesssim$10 $\mu$m &  50--500 & surface &      ppd   & (9)\\
 $-$chemisputter & refr. org.  & $\lesssim$10 $\mu$m &  $>$500 & surface &      ppd   & (10)\\
  $-$chondrule & refr. org.  & $\sim$1 cm &  $>$1200 & midplane &     ppd,dd   & (11)\\
  \hspace{6.5pt}formation &  & &   &  &        &\\
    $-$thermal & refr. org.  & 1-100 km &  $<$1200 & parent body &      dd   & (12) \\
  \hspace{6.5pt}metamorphism &  & &   &  &        &\\
    $-$differentiation & silicate,alloy,  & 10$^{0-4}$ km &  $>$1500& parent body &      dd   & (13)\\
    & vapor.  &  &  &      &  & \\
      $-$degassing & silicate,vapor.  & 10$^{0-4}$ km &  $>$1500& parent body &      dd   & (14)\\
 \hline
\end{tabular}
\end{center}
\begin{tabnote}
$^{\rm a}$ ($+$) implies a gain term and ($-$) signifies loss.\\
$^{\rm b}$surface = active in layers exposed to UV-radiation whether in disk or dense core. deep surface = below UV exposed layers in either disk or dense core and also active at depth. midplane = central layer of disk where dust has settled. parent body = active in planetesimal or planet. plan-emb = active in envelope of planetary embryo. \\
$^{\rm c}$dc = dense core, ppd = protoplanetary disk, dd = debris disks\\
$^{\rm d}$(1) \citet{1989ApJ...341..372F, 1992ApJ...401..269C, Jager09}; (2) \citet{2007ARA&A..45..339B,2009ARA&A..47..427H}; (3) \citet{2024Sci...386..810W, 2023Sci...382.1411Z}; (4) \citet{2012Sci...336..452C, 1999Sci...283.1135B} (5) \citet{2014A&A...570A..32B}; (6) \citet{2022ESC.....6..597M}; (7) \citet{Nakano03, 2017A&A...606A..16G,vanthoff2020, Li21}; (8) \citet{2018A&A...611A..65B, 2023MNRAS.523.6186W} (9) \citet{2014A&A...569A.119A}; (10) O: \citet{Lee10}, OH: \citet{2017A&A...606A..16G}, H:H$_2$:H$_2$O: \citet{2025A&A...694A..89B}; (11) \citet{1998GeCoA..62..903H}; (12) \citet{nakamura2005post}; (13) \citet{Hirschmann16} (14) \citet{2023ASPC..534..907L,Gaillard22,Schlichting15}.\\
$^{\rm e}$ This size of 1~m for sublimation has uncertainty but is estimated for material that has thermalized the entire body to the temperature of the enshrouding H$_2$ gas and small dust.   Comets, of course, sublimate volatiles from bodies with sizes $\gg$1~m and thus there is some ambiguity at to what size means in the context of sublimation.  Models of cometary heating suggest the outer few meters are heated to levels that release  volatile ices \citep{2002EM&P...89...27P}. We adopt 1~m for full body thermalization based on these estimates. However, this is uncertain.
\end{tabnote}
\end{table}

\newpage


\end{document}